\documentclass[conference]{IEEEtran}
\IEEEoverridecommandlockouts
\usepackage{amsmath,amssymb,amsfonts}
\usepackage{algorithmic}
\usepackage{graphicx}
\usepackage{textcomp}
\usepackage{xcolor}

\usepackage[style=ieee, doi=false, isbn=false, url=false]{biblatex}
\AtEveryBibitem{\clearfield{urlyear}}
\DefineBibliographyStrings{english}{
  url = \adddot\addspace Available,
}

\usepackage{multirow}

\usepackage{acro}

\usepackage{xspace}

\usepackage[linesnumbered,ruled,vlined]{algorithm2e}

\usepackage[olditem,oldenum]{paralist}
\usepackage{enumitem}

\usepackage[most]{tcolorbox}

\usepackage{booktabs}

\usepackage{colortbl}

\usepackage{siunitx}

\usepackage{supertabular}

\usepackage{tabularx}

\usepackage{threeparttable}

\usepackage{pifont} %
\usepackage{fontawesome5}
\usepackage{wasysym} %
\usepackage{amsmath} %
\usepackage{graphicx}
\usepackage{hyperref}

\hypersetup{
    colorlinks=true,
    linkcolor=blue,
    filecolor=magenta,      
    urlcolor=cyan!50!black,
    }

\usepackage{xcolor}

\usepackage[all]{nowidow}

\usepackage{listings}

\makeatletter
\let\old@lstKV@SwitchCases\lstKV@SwitchCases
\def\lstKV@SwitchCases#1#2#3{}
\makeatother
\usepackage{lstlinebgrd}
\makeatletter
\let\lstKV@SwitchCases\old@lstKV@SwitchCases

\lst@Key{numbers}{none}{%
    \def\lst@PlaceNumber{\lst@linebgrd}%
    \lstKV@SwitchCases{#1}%
    {none:\\%
     left:\def\lst@PlaceNumber{\llap{\normalfont
                \lst@numberstyle{\thelstnumber}\kern\lst@numbersep}\lst@linebgrd}\\%
     right:\def\lst@PlaceNumber{\rlap{\normalfont
                \kern\linewidth \kern\lst@numbersep
                \lst@numberstyle{\thelstnumber}}\lst@linebgrd}%
    }{\PackageError{Listings}{Numbers #1 unknown}\@ehc}}
\makeatother

\definecolor{dkgreen}{rgb}{0,0.6,0}
\definecolor{gray}{rgb}{0.5,0.5,0.5}
\definecolor{mauve}{rgb}{0.58,0,0.82}

\usepackage{makecell}

\definecolor{sangria}{rgb}{0.57, 0.0, 0.04}
\definecolor{pinegreen}{rgb}{0.0, 0.47, 0.44}
\definecolor{rossocorsa}{rgb}{0.83, 0.0, 0.0}
\definecolor{ao}{rgb}{0.0, 0.0, 1.0}
\definecolor{deepjunglegreen}{rgb}{0.0, 0.29, 0.29}
\definecolor{dartmouthgreen}{rgb}{0.05, 0.5, 0.06}

\definecolor{tableblue}{HTML}{E7F0FA}   %
\definecolor{tablegray}{HTML}{F4F4F4}   %
\definecolor{tableorange}{HTML}{FEF2E5} %

\newcommand\CC{C\nolinebreak[4]\hspace{-.05em}\raisebox{.4ex}{\relsize{-2}{\textbf{++}}}}

\newcommand{\tool}{\textsc{Anota}\xspace}
\newcommand{\toolfuzz}{\textsc{Anota+Fuzzer}\xspace}

\newcommand{\numzerodays}{22\xspace}
\newcommand{\numcves}{17\xspace}
\newcommand{\numvlunids}{20\xspace}
\newcommand{\numreproducedbugs}{43\xspace}
\newcommand{\numreproducedapplications}{47\xspace}

\def\7zip{\texttt{7zip}\xspace}

\def\e2fsck{\texttt{e2fsck}\xspace}

\def\mke2fs{\texttt{mke2fs}\xspace}

\newcommand{\cmark}{\textcolor{pinegreen}{\ding{51}}\xspace}
\newcommand{\xmark}{\textcolor{rossocorsa}{\ding{56}}\xspace}

\newcommand*\circled[1]{\tikz[baseline=(char.base)]{
            \node[shape=circle,draw,inner sep=.4pt] (char) {#1};}}

\newcommand*\blackcircled[1]{\tikz[baseline=(char.base)]{
            \node[shape=circle,fill,inner sep=.4pt] (char) {\textcolor{white}{#1}};}}

\newcommand{\result}[1]{%
\begin{tcolorbox}[colback=blue!4!white,colframe=black,boxrule=0.6pt,arc=5pt,
      left=4pt,right=4pt,top=4pt,bottom=4pt,boxsep=0pt,width=\columnwidth]%
      {#1}
\end{tcolorbox}%
}

\DeclareAcronym{asan}{
    long={AddressSanitizer},
    short={ASan},
    cite={serebryany_addresssanitizer_2012}
}
\DeclareAcronym{ubsan}{
    long={UndefinedBehaviorSanitizer},
    short={UBSan},
    cite={undefinedbehaviorsanitizer_2013}
}
\DeclareAcronym{msan}{
    long={MemorySanitizer},
    short={MSan},
    cite={stepanov_memorysanitizer_2015}
}
\DeclareAcronym{put}{
    long={program under test},
    short={PUT},
}
\DeclareAcronym{ssrf}{
    long={Server-Side Request Forgery},
    short={SSRF},
}
\DeclareAcronym{xss}{
    long={Cross-Site Scripting},
    short={XSS},
}
\DeclareAcronym{toctou}{
    long={time-of-check to time-of-use},
    short={TOCTOU},
}
\DeclareAcronym{ebpf}{
    long={Extended Berkeley Packet Filter},
    short={eBPF},
    cite={ebpf}
}
\DeclareAcronym{dbi}{
    long={Dynamic Binary Instrumentation},
    short={DBI}
}
\DeclareAcronym{llm}{
    long={Large Language Model},
    short={LLM}
}

\begin{document}

\title{Anota: Identifying Business Logic Vulnerabilities via Annotation-Based Sanitization}

\author{
    \IEEEauthorblockN{
    Meng Wang\IEEEauthorrefmark{1},
    Philipp Görz\IEEEauthorrefmark{1},
    Joschua Schilling\IEEEauthorrefmark{1},
    Keno Hassler\IEEEauthorrefmark{1},
    Liwei Guo\IEEEauthorrefmark{2},
    Thorsten Holz \IEEEauthorrefmark{3},
    Ali Abbasi\IEEEauthorrefmark{4}
                      }
    \IEEEauthorblockA{
        \IEEEauthorrefmark{1}\IEEEauthorrefmark{4}CISPA Helmholtz Center for Information Security, 
        \IEEEauthorrefmark{2}University of Electronic Science and Technology,\\
        \IEEEauthorrefmark{3}Max Planck Institute for Security and Privacy
    }
}

\IEEEoverridecommandlockouts
\makeatletter\def\@IEEEpubidpullup{6.5\baselineskip}\makeatother
\IEEEpubid{\parbox{\columnwidth}{
		Network and Distributed System Security (NDSS) Symposium 2026\\
		23 - 27 February 2026 , San Diego, CA, USA\\
		ISBN 979-8-9919276-8-0\\  
		https://dx.doi.org/10.14722/ndss.2026.[23$|$24]xxxx\\
		www.ndss-symposium.org
}
\hspace{\columnsep}\makebox[\columnwidth]{}}

\maketitle
\begin{abstract}
Detecting business logic vulnerabilities is a critical challenge in software security. These flaws come from mistakes in an application's design or implementation and allow attackers to trigger unintended application behavior. %
Traditional fuzzing sanitizers for dynamic analysis excel at finding vulnerabilities related to memory safety violations
but largely fail to detect business logic vulnerabilities, as these flaws require understanding application-specific semantic context.
Recent attempts to infer this context, due to their reliance on heuristics and non-portable language features, are inherently brittle and incomplete.
As business logic vulnerabilities constitute a majority (27 of the CWE Top 40) of the most dangerous software weaknesses in practice, this is a worrying blind spot of existing tools.

In this paper, we tackle this challenge with \tool, a novel human-in-the-loop sanitizer framework. 
\tool introduces a lightweight, user-friendly annotation system that enables users to directly encode their domain-specific knowledge as lightweight annotations that define an application's intended behavior.
A runtime execution monitor then observes program behavior, comparing it against the policies defined by the annotations, thereby identifying deviations that indicate vulnerabilities.
To evaluate the effectiveness of \tool, we combine \tool with a state-of-the-art fuzzer and compare it against other popular bug finding methods compatible with the same targets.
The results show that \toolfuzz outperforms them in terms of effectiveness.
More specifically, \toolfuzz can successfully reproduce \numreproducedbugs known vulnerabilities, and discovered \numzerodays previously unknown vulnerabilities (\numcves CVEs assigned) during the evaluation.
These results demonstrate that \tool provides a practical and effective approach for uncovering complex business logic flaws often missed by traditional security testing techniques.

\end{abstract}

\section{Introduction}

\label{sec:intro}
Fuzzing, a dynamic program analysis technique, is widely used to discover software vulnerabilities across diverse systems. Fuzzing operates by supplying the Program Under Test (PUT) with a large number of malformed or unexpected inputs, aiming to trigger vulnerable states.
However, the fuzzer itself is typically oblivious to whether a given input has successfully exposed a vulnerability. 
This is where sanitizers play an indispensable role. Sanitizers are instrumental tools that monitor program execution, detect predefined classes of errors, and report these violations to the fuzzer, thereby locating the vulnerability-inducing inputs~\cite{sanitization-sok}. 
Historically, sanitizers are developed for detecting memory-related~\cite{serebryany_addresssanitizer_2012, stepanov_memorysanitizer_2015,chen2018iotfuzzingmemorycorr,muench2018youfuzzingmemorycorr,drmemory, valgrind}, undefined behavior~\cite{undefinedbehaviorsanitizer_2013}, and data race~\cite{serebryany2009threadsanitizer, helgrindplus, binarysan} vulnerabilities in memory-unsafe languages such as C and \CC. 
More recently, adapting this technique to memory-safe languages such as Python and PHP has led to the development of sanitizers targeting a broader range of vulnerabilities.
These include issues such as command injection, Server-Side Request Forgery (SSRF), and path traversal~\cite{witcher, cefuzz, Atropos, SSRFuzz, edefuzz, ppfuzz, ufuzzer, webfuzz, csrffuzz, infoflowfuzz}.

To examine the extend to which these recent sanitizers are capable of detecting the most prevalent vulnerabilities, we analyzed the 40 most dangerous software weaknesses (CWE Top 40)~\cite{mitreMostDangerousTop40}.Despite recent progress in sanitizer development, we find that a considerable number of these critical vulnerabilities are either not addressed at all or inadequately addressed. This worrying observation forms the primary motivation for our work: there is a need to develop sanitizers to light up this detection blind spot. 
As shown in Table~\ref{tab:table_cwe}, our analysis of fuzzing sanitizers identifies two groups of weaknesses: \textit{Unaddressed Weaknesses} and \textit{Narrowly-Addressed Weaknesses}. 
\textbf{Unaddressed Weaknesses} (blue rows in Table~\ref{tab:table_cwe}): refers to weaknesses for which no sanitizer has been proposed yet. A prominent example is authorization: the sanitizer needs to know the application's context to identify whether the user has the privilege to access certain resources like reading/writing variables, calling functions, or executing privileged code. 
Similar obstacles block other CWE entries, such as Uncontrolled Search Path Element, for which the sanitizer needs to have knowledge of the developer's intended file system access privileges. 

\textbf{Narrowly-Addressed Weaknesses} (grey rows in Table~\ref{tab:table_cwe}): 
Even when sanitizers do exist for certain vulnerabilities, their solutions often have limited generalizability. Many established approaches~\cite{Atropos, ufuzzer, ODDFuzz, URadar} exhibit language-specificity, depending on inferring program context from idiosyncratic language-specific features, thereby limiting their applicability to a single language, without offering paths to generalize the approach. Other techniques~\cite{edefuzz, infoflowfuzz, he2020ct} impose strong pre-conditions that are often unattainable in practice, such as EDEFuzz~\cite{edefuzz}, which requires a rendered web interface for its operation, a setup not always available for every application.

Our investigation into these detection gaps reveals a common cause: these weaknesses are overwhelmingly business logic vulnerabilities.
Detecting such flaws requires a deep understanding of an application's specific rules and intended workflows: a level of understanding semantic context that automated tools struggle to achieve.
To pursue full automation, existing tools are forced to approximate this understanding using predefined heuristic patterns of behavior that seem suspicious.
However, these heuristics are selected artificially and cannot capture the nuanced, application-specific context required to reliably distinguish legitimate behavior from a true security violation.

Consider the example of a file upload feature.
State-of-the-art tools like Atropos~\cite{Atropos} attempt to detect business logic vulnerabilities like unrestricted file uploads using narrow, hard-coded heuristics.
More specifically, Atropos instruments a predefined list of standard PHP functions and flags only PHP file uploads as malicious.
This detection logic is inherently brittle and is easily bypassed if an attacker uploads a different dangerous file type or if the application uses a custom function not on the tool's predefined list.
This reliance on superficial patterns fails against other context-dependent attacks, such as an authenticated user exploiting a path traversal flaw to upload a file into another user's private folder.
Detecting such a violation requires specific contextual knowledge that the heuristic lacks:
Who is the requesting user? What are their permissions for the target directory?
For a fully automated tool, reliably inferring these fine-grained policies is fundamentally challenging;
The pursuit of automation often comes at the expense of generalizability and accuracy.
However, such semantic knowledge is readily available to an application's developer or user, who can easily define the intended policy.

In the research area of vulnerabilities in memory-unsafe programming languages, there has been a shift from early attempts to find memory errors solely by observing program crashes~\cite{miller1990empirical} towards sanitization.
Tools like AddressSanitizer~\cite{serebryany_addresssanitizer_2012} detect the violation at its source (e.g., an illegal memory operation), thereby finding bugs that would not have manifested in observable crashes.
This works well because the correct behavior is \emph{implicitly} encoded for these vulnerabilities---a memory violation is never correct.
This tool has been instrumental in wide-scale fuzz testing of open-source software, uncovering more than 36,000 security bugs~\cite{google2022clusterfuzz}.
We would like to benefit similarly from sanitization in the field of business-logic vulnerabilities.
However, in this field, the correct behavior is not clear and has to be stated \emph{explicitly} to overcome the limitations of narrow heuristics.

To this end, we introduce \tool, a novel human-in-the-loop sanitizer framework built with the classic systems principle~\cite{10.1145/1067629.806531} of separating policy from mechanism. Our approach empowers developers to encode their knowledge, avoiding the pitfalls of automated context inference.
\tool is designed to work with any existing fuzzer, typically by reporting policy violations as program crashes.
\tool consists of two core components: A lightweight, easy-to-use annotation system through which a developer or security analyst defines application-specific semantic rules (the policy),
and a general-purpose, language-agnostic runtime policy monitor that enforces these policies during the program execution (the mechanism).
This design fundamentally shifts the generalization effort.
Instead of building increasingly complex and brittle heuristics, \tool provides an extensible framework for users to express their semantic understanding of the application's intended behavior.
The human-in-the-loop methodology has proven highly effective in other challenging domains such as software verification and interactive machine learning~\cite{bertot2013interactive, shoshitaishvili2017rise, wu2022survey, mosqueira2023human}. 
The underlying idea is that humans, equipped with deep contextual understanding and knowledge of the system's business logic, can provide invaluable input when facilitated by an intuitive and effective generic API.

We implemented a prototype of \tool for Python applications, chosen for its widespread popularity~\cite{tiobeTIOBEIndex} and the prevalence of complex applications built with it.
To evaluate the effectiveness and overhead of \tool, we integrate it with the Python fuzzer Atheris~\cite{atheris} to create \toolfuzz.
We evaluate \toolfuzz against an unmodified Atheris fuzzer, showing \tool could empower the baseline fuzzer to detect various business-logic vulnerabilities and achieve significantly higher precision and recall on our benchmark datasets.
Furthermore, we validated \tool's practical utility by re-discovering \numreproducedbugs known vulnerabilities and uncovering \numzerodays new vulnerabilities in popular, actively maintained Python projects.
To date, \numcves of these discoveries have been assigned CVE identifiers.
An annotation study further confirmed that \tool's annotations are intuitive, and our performance evaluations show that it incurs minimal runtime overhead (about 10\% in an artificial worst-case setup) while benchmarking using Python Performance Benchmark Suite~\cite{Pythonperfbench}.

\smallskip
\noindent
\textbf{Contributions.} We make the following key contributions:
\begin{compactitem}
\item We systematically analyze existing sanitizers against the CWE Top 40, demonstrating that the most critical unaddressed weaknesses are business logic vulnerabilities whose detection requires deep semantic context that current fuzzing sanitizers cannot capture.
\item We propose a new sanitization paradigm based on the principle of separation of policy and mechanism. We design a lightweight annotation system that enables developers to formally express a program's intended behavior, thereby guiding dynamic analysis.
\item We present the design and implementation of \tool, a sanitizer that instantiates our paradigm by integrating developer annotations with a runtime monitor to detect policy violations at their source.
\item We implement \tool for Python and create \toolfuzz. Through comprehensive evaluation, we demonstrate \tool's effectiveness in finding real-world bugs, uncovering \numzerodays impactful zero-day vulnerabilities, and outperforming state-of-the-art tools.
\end{compactitem}

Following our commitment to open science, we make our implementation and evaluation scripts available online at \url{https://github.com/ANOTA-Sanitizer/ANOTA}.

\section{Background}
\label{sec:background}

We start by providing an overview of sanitizers developed to detect vulnerabilities listed in the CWE Top 40, then analyze the reason why certain vulnerabilities are Unaddressed or Narrowly-Addressed.  
We conclude with a motivating example to illustrate our approach. 

\subsection{Limitations of Existing Sanitizers} 
Sanitizers are a key component of dynamic analysis like fuzzing. They act as runtime oracles that monitor a program's execution to determine if its behavior indicates a security vulnerability. Sanitizers have evolved significantly. Early approaches often relied on simple signals like program crashes~\cite{miller1990empirical}  to detect bugs. In contrast, modern tools like AddressSanitizer~\cite{serebryany_addresssanitizer_2012} are far more sophisticated, excelling at finding errors like memory corruption that violate universal rules of program execution.

However, this success does not fully extend to all the vulnerabilities prevalent in modern applications. Our analysis of sanitization capabilities against the CWE Top 40 vulnerabilities (detailed in Table~\ref{tab:table_cwe}) reveals a blind spot. This blind spot is most apparent for what we term \textit{Unaddressed Weaknesses}. Many critical vulnerabilities, particularly those related to authorization and authentication, are invisible to existing sanitizers because detecting them requires application-specific context. For instance, to find an improper privilege management flaw, a sanitizer must understand the application's intended permission model to know if a user's access attempt is unauthorized—a level of semantic knowledge that generic tools lack. 

In other cases, where sanitizers do exist, they are often \textit{Narrowly-Addressed Weaknesses}. These tools attempt to infer context through methods that are brittle and impractical. Many rely on language-specific heuristics; for example, Atropos~\cite{Atropos} and PHUZZ~\cite{PHUZZ} are tied to standard PHP functions. Atropos~\cite{Atropos} also pre-defined a conservative heuristic as mentioned in the previous section. Neither method is portable, but both can be easily bypassed. ODDFuzz~\cite{ODDFuzz} illustrates this tight coupling with its design specifically tailored to sensitive call-sites within Java applications. Furthermore, other tools depend on restrictive operational pre-conditions, such as the GUI required by EDEFuzz~\cite{edefuzz} or the identical execution environments and significant manual setup needed by FLOWFUZZ~\cite{infoflowfuzz}. 

All of these limitations stem from a single, fundamental challenge: the difficulty of automatically inferring deep, application-specific context. The repeated failure of automated heuristics to reliably capture this context motivates the need for an alternative approach—one that can directly equip sanitizers with the crucial awareness they currently lack. We propose that an intuitive and lightweight annotation framework, enabling users to systematically embed their domain-specific knowledge and security intent directly into the program, can grant sanitizers access to the precise contextual awareness they currently lack. Such an approach would circumvent the limitations of language-dependent heuristics and the complexities of inferring context from external interfaces, thereby paving the way for more precise, effective, and broadly applicable vulnerability detection.

\subsection{Motivating Example} 
\label{sec:motivation}

Consider the Python program in Listing~\ref{lst:motivation_example}.
The developer wants to filter out certain network schemes and host names to avoid fetching content from remote user-provided URLs with such patterns.
The developer imports the \texttt{urllib} library from Python's standard library for working with URLs.
The developer defines schemes and host names to be blocked (lines 2--3) and relies on the \texttt{urlparse} API to parse the input URLs (lines 4--5).
Then the code checks if the URL matches the block pattern (lines 6--9) to decide whether to retrieve the remote sites' contents (lines 10--11).

\begin{lstlisting}[float, captionpos=b, language=Python, caption={Motivation example for our approach.}, label={lst:motivation_example}]
 def SafeURLOpener(input_link):
    block_schemes = ["file", "php", "ftp", "data"]
    block_host = ["youtube.com", "instagram.com"]
    input_scheme = urlparse(input_link).scheme
    input_hostname = urlparse(input_link).hostname
    if input_scheme in block_schemes:
        return
    if input_hostname in block_host:
        return
    target = urllib.request.urlopen(input_link)
    print(target.read())
\end{lstlisting}

Although at first glance this implementation looks acceptable, it contains the vulnerability \href{https://www.cve.org/CVERecord?id=CVE-2023-24329}{\texttt{CVE-2023-24329}}, introduced by \texttt{urllib} in its API \texttt{urllib.parse.urlparse}.
This API mistakenly parses URLs beginning with whitespace, which allows an attacker to bypass the filtering mechanism shown in Listing~\ref{lst:motivation_example}. 
For instance, if the remote user provides the URL \texttt{``https://youtube.com''}, the \texttt{SafeURLOpener} will block the request. 
However, if the URL is prefixed with whitespace (e.g., \texttt{``\textvisiblespace\textvisiblespace\textvisiblespace https://youtube.com''}), the blocking mechanism will be bypassed. %

Detecting this vulnerability is challenging because it requires understanding application-specific business logic, not just common bug heuristics or patterns %
Static analysis fails, while it can identify the urlopen API call, it is blind to the custom filtering logic that defines the intended policy. Dynamic analysis like fuzzing also fails; although it can possibly trigger the fault, the violation goes undetected because it neither crashes the program nor violates a rule known to existing sanitizers.
Ultimately, both approaches have the same core problem of lacking access to the application's context. Effective detection is only possible by comparing a program's runtime behavior against its application-specific semantic
rules (the policy) to tell the difference between a flaw and correct execution.

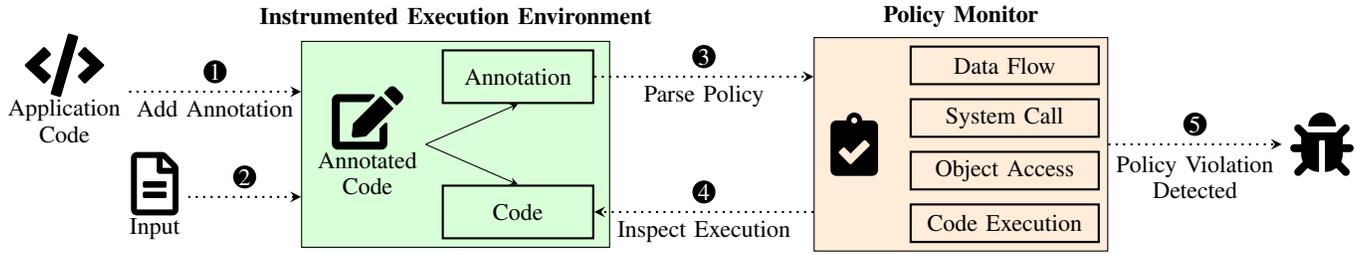
\begin{figure*}
  \small
  \centering
  \usetikzlibrary{fit}

  \pgfdeclarelayer{background}
  \pgfdeclarelayer{foreground}
  \pgfsetlayers{background, main, foreground}

  \tikzstyle{polbox} = [rectangle, minimum width=2.5cm, minimum height=0.5cm, text centered, draw=black, thick]
  \tikzstyle{envbox} = [rectangle, minimum width=2cm, minimum height=0.7cm, text centered, draw=black, thick]
  \tikzstyle{arrow} = [thick,->,>=stealth, dotted]
  \tikzstyle{internal-arrow} = [->,>=stealth]

  \begin{tikzpicture}[node distance=2cm]
    \node (code-ann) [] {\makecell[c]{{\Huge\faEdit}\\Annotated\\Code}};
    \node (code) [envbox, right of=code-ann, yshift=-0.9cm] {Code};
    \node (ann) [envbox, right of=code-ann, yshift=0.9cm] {Annotation};
    \node (dummy) [right of=code-ann, xshift=1cm] {};

    \draw[internal-arrow] (code-ann.east) -- (ann.south);
    \draw[internal-arrow] (code-ann.east) -- (code.north);

    \begin{pgfonlayer}{background}
      \node (env-box) [draw=black, fill=green!15,fit=(code-ann) (code) (ann) (dummy)] {};
    \end{pgfonlayer}
    \node (env-text) [above of=env-box, yshift=-0.3cm] {\textbf{Instrumented Execution Environment}};

    \node (app-code) [left of=env-box, xshift=-3.2cm, yshift=0.7cm] {\makecell[c]{{\Huge\faCode}\\Application\\Code}};
    \draw[arrow] (app-code) -- (env-box.west|-app-code) node[midway, above, solid] {\blackcircled{1}} node [midway, below] {Add Annotation};

    \node (input) [left of=env-box, xshift=-2cm, yshift=-0.7cm] {\makecell[c]{{\Huge\faFile*[regular]}\\Input}};
    \draw[arrow] (input) -- (env-box.west|-input) node[midway, above, solid] {\blackcircled{2}};
    
    \node (polmon) [right of=env-box, xshift=3.3cm] {\Huge\faClipboardCheck};
    \node (rule1) [polbox, right of=polmon, yshift=1.05cm] {Data Flow};
    \node (rule2) [polbox, right of=polmon, yshift=0.35cm] {System Call};
    \node (rule3) [polbox, right of=polmon, yshift=-0.35cm] {Object Access};
    \node (rule4) [polbox, right of=polmon, yshift=-1.05cm] {Code Execution};

    \begin{pgfonlayer}{background}
      \node (polmon-box) [draw=black, fill=orange!15,fit=(polmon) (rule1) (rule2) (rule3) (rule4)] {};
    \end{pgfonlayer}
    \node (polmon-text) [above of=polmon-box, yshift=-0.3cm] {\textbf{Policy Monitor}};

    \draw[arrow] (ann.east) -- (polmon-box.west|-ann) node[midway, above, solid] {\blackcircled{3}} node [midway, below] {Parse Policy};
    \draw[arrow] (polmon-box.west|-code) -- (code.east) node[midway, above, solid] {\blackcircled{4}} node [midway, below] {Inspect Execution};

    \node (bug) [right of=polmon-box, xshift=2.8cm] {\Huge\faBug};
    \draw[arrow] (polmon-box) -- (bug) node[midway, above, solid] {\blackcircled{5}} node [midway, below] {\makecell[c]{Policy Violation\\Detected}};

  \end{tikzpicture}

  \caption{High-level overview of \tool.}
  \label{fig:overview}
\end{figure*}

\section{Design}
\label{sec:design}
In the following, we present the design and implementation of \tool, a human-in-the-loop framework that enables a developer or security analyst to sanitize hard-to-detect business logic bugs using annotations.
A high-level overview of \tool is shown in Figure~\ref{fig:overview}.
At its core, a policy monitoring module escalates policy violations to a crash that can be detected by fuzzing. 
Our approach consists of two phases: an initial annotation phase and an application testing phase.
In the first phase, the user of \tool adds annotations to the code in step \blackcircled{1}. 
In the testing phase, a dynamic analysis tool like a fuzzer executes \blackcircled{2} the annotated application with a series of test inputs.
The instrumented execution environment parses the policies reflected in the annotations \blackcircled{3} and the policy monitor enforces the policies and checks the execution of the program \blackcircled{4}.
If a policy violation is detected during the testing process, it is included in the vulnerability report \blackcircled{5}. 
In the following, we present our design of the annotations and the policy monitor in more detail.

\subsection{Annotations}
The core design principle of the annotation system is to enable developers and security analysts to directly encode their contextual understanding of the program into a set of enforceable runtime policies. 
These policies are defined through annotations that grant or revoke permissions for program actions, offering fine-grained control, allowing developers to define policies for specific parts of an application using two models: a blocklist, which prohibits a defined set of actions while allowing all others, or a more restrictive allowlist, which permits only a defined set of actions while denying all others by default. Also, annotations can be controlled individually and toggled during runtime.

The annotations provided by \tool (summarized in Table~\ref{Tab:annotation_overview}) are based on the runtime behavior caused by the CWE Top 40 vulnerabilities~\cite{mitreMostDangerousTop40}. Our analysis of business logic vulnerabilities detailed in Table~\ref{tab:table_cwe}, identified four critical types of deviant behavior: (1) unintended system call usage, (2) unintended code execution, (3) unintended variable access, and (4) unintended data flow. Consequently, we designed four primary annotation types to counter these behaviors, along with a fifth to disable annotations during runtime 

As shown in Table~\ref{tab:table_cwe}, our set of annotations was designed to be expressive enough to address the business logic vulnerabilities found in the CWE Top 40.
The system is also extensible by design.
Because we separate the annotation definitions (policy) from the runtime monitor (mechanism), users can add new annotation types for other vulnerabilities without re-engineering the core system.
This typically just requires defining the new annotation's syntax and implementing a corresponding check in the runtime monitor, allowing the framework's detection capabilities to expand as new vulnerability classes emerge.

\begin{table}[t]\centering
  \caption{Overview of Annotations Provided by \tool} 
  \label{Tab:annotation_overview}
  \scriptsize
  \begin{tabularx}{\linewidth}{@{ }l@{ \ }l@{ \ }c@{ }X@{ }}
    \toprule
    {\bf Type}       & {\bf Options}        & \multicolumn{2}{l}{\bf Example} \\
    \midrule
    System Call      & Syscalls and args    & \circled{1} & \texttt{SYSCALL.BLOCK('execv', 'execveat', 'execve')} \\
                     &                      & \circled{2} & \texttt{SYSCALL.READ.BLOCK (PATH='/etc/')} \\
                     &                      & \circled{3} & \texttt{SYSCALL.EXECVE.ALLOW (PATH='/bin/ls')} \\
    Data Flow        & Sanitization, Sink   & \circled{4} & \texttt{TAINT(pwd, sanitization= [hash], Sink=[print])} \\
    Object Access    & Read, Write, Execute & \circled{5} & \texttt{WATCH.ALLOW(admin\_data,'r')} \\
                     &                      & \circled{6} & \texttt{WATCH.BLOCK(admin\_api,'x')} \\
                     & & \circled{7} & \texttt{WATCH.CON(passwd\_cmp)} \\

    Code Execution   & Condition            & \circled{8} & \texttt{EXECUTION.BLOCK (user.type!='admin')} \\ 
    Annotation Clear & Annotation           & \circled{9} & \texttt{CLEAR(SYSCALL.EXECVE. ALLOW(PATH='ls'))} \\
                     &                      & \circled{10} & \texttt{SYSCALL.NETWORK.CLEAR()} \\
    \bottomrule
  \end{tabularx}
\end{table}

\textbf{System Call} annotations enforce access only to the allowed system calls and resources.
Using \texttt{SYSCALL. [ALLOW|BLOCK](List of System Calls)} specifies which system call is allowed or blocked.
For example, \circled{1} in Table~\ref{Tab:annotation_overview} blocks three \texttt{execv}-related system calls.
Additionally, more fine-grained policies can restrict specified system calls to specific resources using the syntax:
\texttt{SYSCALL.[SYSCALL NAME|SYSCALL CLASS].[ALLOW|BLOCK](arguments)}.
As shown in \circled{3}, this allows a user to define that the \texttt{/etc/} directory should not be accessible by the \texttt{read} system call in \circled{2}, or the \texttt{execve} system call can only use the \texttt{/bin/ls} executable.

To make policy writing more intuitive for users, \tool provides a simplified syntax (``syntactic sugar'') for common system calls.
For example, \texttt{FILE} can be used to control file/directory, attribute modification, and read/write permissions for all file-accessing system calls at once.
Similarly, \texttt{NETWORK} represents all network-related system calls, and options like \texttt{SCHEME} are provided to control the URL scheme.
Note that a more knowledgeable user could also directly define raw policies to control each argument of a system call for even more fine-grained control.
Additionally, wildcards can be used to define matches similarly to shell expansion, e.g., \texttt{SYSCALL.READ.BLOCK(PATH='/upload\_folder/*. php')}
blocks read access to all PHP scripts in \texttt{upload\_folder}.

\textbf{Data Flow} annotations instruct \tool to track the data flow of a variable.
The \texttt{TAINT(Target\_Variable, Sanitiza\-tion Method, Sink)} annotation specifies which variable(s) to taint.
We provide two optional arguments for this annotation:
An option to clear the taint if the variable is used as the argument to the given sanitization function and an option to mark sink functions that are prohibited from receiving the tainted variable as an argument.
The data flow annotation supports tainting various objects: If the object is a variable, the variable itself will be tainted, and if the object is a callable object such as a function, objects returned from the callable object will always be tainted.
In example \circled{4}, \texttt{pwd} is disallowed to reach the \texttt{print} function, except if it is passed to \texttt{hash} first.
By default, the sink is \texttt{write()}, which can output data to any untrusted domain (e.g., file system or network).

\textbf{Object Access} can be controlled by two different types of annotations.
With \texttt{WATCH.[ALLOW|BLOCK](Target Object, Permission)} privilege can be assigned. There are three different types of privileges:
A read permission (\texttt{r}) grants the program the ability to read the value of an object, including assigning it to other variables or passing it to other modules; an example can be seen in \circled{5}.
Similarly, write permission (\texttt{w}) grants the program the ability to modify the value of an object and includes the ability to remove the object.
Lastly, an execute permission (\texttt{x}) grants the program the ability to execute an object, providing that the object is a callable like a function or a method, as shown in example \circled{6}.
The second type of annotation syntax is \circled{7} \texttt{WATCH.CON(Target Object)} which means the annotated object's execution time should be statistically constant. 

\textbf{Code Execution} annotations can be applied to code regions that should not be allowed to be executed, unless a condition is met. 
Example \circled{8} forbids execution if \texttt{user.type} is not \texttt{'admin'}.

\textbf{Annotation Clearing} is used to remove a specific annotation policy or all annotation policies if there are no arguments given.
Example \circled{9} disables the specific policy that limits the \texttt{execve} system call to the \texttt{ls} executable.
Example \circled{10} clears policies on all network-related system calls.
This type of annotation can be helpful if the user wants to apply different policies for different parts of the code.

\subsection{Policy Monitor}
The runtime policy monitor is responsible for translating developer annotations into enforceable runtime policies and identifying the unintended behaviors of the applications as policy violations. 
To detect such violations, our monitor instruments the program at policy-relevant locations (e.g., object access or system call sites). When an instrumented point is reached during execution, the monitor intercepts the event, evaluates the runtime context against the specified policy, and reports any discrepancy as a security flaw. The following sections detail this instrumentation and enforcement mechanism.

\subsubsection{Data Flow Monitor}
To detect vulnerabilities related to data flow issues like sensitive information exposure, our approach relies on dynamic data flow analysis~\cite{schwartz2010SoK}. 
A tracking tag (\emph{taint}) is added to the variable annotated by the user, and the data flow monitor will track the data flow and control the taint propagation through the program execution based on the annotation arguments.

As shown in Listing~\ref{lst:data_flow_san_annotated}, the developer mistakenly outputs the sensitive variable \texttt{user\_credential} into the local log file.
By adding a \texttt{Data Flow} annotation (line 2), the user of \tool can ask the data flow monitor to track the variable \texttt{user\_credential} in the program execution and remove the taint when the variable is sanitized. 
Then the policy violation is detected (line 4) as \texttt{user\_credential} is being written into a local file, and \tool will identify the unintended behavior.
Hence, the policy monitor is notified that a new vulnerability has been detected.
Writing \texttt{hashed\_credential}, however, would not trigger the violation because it has been sanitized by the sanitization function \texttt{hash} defined in the annotation.

\begin{lstlisting}[float, caption={Dataflow example in PHP.}, label={lst:data_flow_san_annotated}, linebackgroundcolor={\ifnum \value{lstnumber} = 2 \color{green!20} \fi }]
$user_credential = get_credential($user_id);
TAINT($user_credential, sanitization=[hash]);
$hashed_credential = hash($user_credential);
$log_message = sprintf("credential %
file_put_contents("logfile.txt", $log_message, FILE_APPEND);
\end{lstlisting}

\subsubsection{System Call Monitor}
To identify vulnerabilities associated with unintended system call usage such as path traversal and unrestricted upload of files, the system call monitor module observes the usage, arguments, and return values of system calls.
Suppose, for instance, that a user writes annotations to confine read access to specific file paths.
To validate whether a \texttt{read} system call violates this policy, the path must be extracted from the file descriptor argument of the \texttt{read} system call.
This argument corresponds to the return value of the corresponding \texttt{openat} system call, and the \texttt{openat} system call's arguments contain the corresponding file path information.
This information is used to determine the file path from which the system call reads data.

At the start of program execution, the system call monitor module attaches itself to the execution process and seamlessly extends its monitoring to subsequent processes spawned by the program.
During program execution, the system call monitor inspects system calls to find if any policy violation occurred.
If a violation occurs---for instance, an unintended system call is invoked or the program writes data to an unintended location---the system call monitor sends a signal to the execution environment.
As a result, the instrumented execution environment could identify that the application is in a vulnerable state.

\begin{lstlisting}[float, caption={Annotated motivation example.}, label={lst:motivation_example_annotated}, linebackgroundcolor={\ifnum \value{lstnumber} = 2 \color{green!20} \fi \ifnum \value{lstnumber} = 3  \color{green!20} \fi \ifnum \value{lstnumber} = 7 \color{green!20} \fi}]  
def SafeURLOpener(inputLink):
    SYSCALL.NETWORK.BLOCK.SCHEME("file","php"...)
    SYSCALL.NETWORK.BLOCK.HOST("youtube.com"...)
    target = urllib.request.urlopen(inputLink)
    print(target.read())
    SYSCALL.NETWORK.CLEAR()
\end{lstlisting}

For the motivating example explained in Section~\ref{sec:motivation}, our approach can successfully detect this vulnerability even if the developer of the code in Listing~\ref{lst:motivation_example} has no knowledge of the underlying faulty library implementation by adding a few system call annotations, as shown in Listing~\ref{lst:motivation_example_annotated}.
Note that our annotation system enables developers to transform their intuition about which URLs should and should not be filtered into an explicit and monitored security policy via a small set of annotations.
\begin{lstlisting}[float, caption={System call example in JavaScript.}, label={lst:syscall_san}, linebackgroundcolor={\ifnum \value{lstnumber} = 4 \color{green!20} \fi}]
app.get('/static/:filename', (req, res) => {
    const file = req.params.filename;
    const staticDir = path.join(cwd, 'static');
    SYSCALL.FILE.ALLOW(staticDir);
    res.sendFile(path.join(staticDir, file));}
\end{lstlisting}

As an additional example, consider Listing~\ref{lst:syscall_san}, where we show an example system call annotation used to detect a path traversal vulnerability.
The developer intends to send a file stored in a folder named \texttt{static} in the current working directory to the remote user based on the request URL.
An annotation is added in line 4 to ensure that file-related system calls can only access the path specified in the argument.

\subsubsection{Access Control Monitor}
\label{sec:access-control-monitor}
Access control related vulnerabilities, like missing authorization or improper authentication, could be exploited in several ways. 
For example, executing privileged code logic, reading/writing privileged variables, or interacting with privileged system resources like files and network resources.
The developer can mark a section of code as privileged with the \texttt{Code Execution} annotation, or a variable as only accessible to the developer with \texttt{Object Access} annotations.
Inside the instrumented execution environment, \tool monitors the code execution status by collecting variable metadata from the stack frame and heap. 
Then it checks access to variables by watching %
read and write operations to that variable.
Similar to previous policies, the policy monitor will be notified when the policy is violated. 
As shown in the example in Listing~\ref{lst:access_control_san_annotated},
only the administrator should be able to add or remove users.
\begin{lstlisting}[float, caption={Access control example in Python.}, label={lst:access_control_san_annotated}, linebackgroundcolor={\ifnum \value{lstnumber} = 4 \color{green!20} \fi }]
if is_auth:
    EXECUTION.BLOCK()
    user_list.append(new_user)
WATCH.ALLOW(user_list, "r")
user_list.remove(existing_user) if is_auth else print("not authenticated")
\end{lstlisting}

\section{Implementation}
\label{sec:implementation}
We implement our design as a prototype named \tool based on CPython (the original Python interpreter), LLVM~\cite{LLVM:CGO04}, and eBPF~\cite{ebpf}.
The prototype implementation consists of about 5,500 lines of Python, C, and Rust code. 

The implementation mirrors the idea of separating policy from mechanism, consisting of two primary components: an Annotation Frontend and a Policy Monitor. The Annotation Frontend is integrated into a modified CPython interpreter to parse annotations and prepare the corresponding security policies. The Policy Monitor enforces these policies and implements the bug sanitizer feedback of the annotations by monitoring program execution.

However, our implementation does not stop at the Python boundary; \tool also supports native extension modules written in C/C++. While the Python interpreter cannot inspect the internal operations of this native code, our policy monitor uses different backends depending on the context: (i) For pure Python code, policy enforcement is handled by hooks within our modified CPython interpreter. (ii) For native C/C++ extensions, we implemented an LLVM instrumentation pass to enable annotation support and policy monitoring for these modules. (iii) For system-level interactions (e.g., file system access), we use an eBPF-based~\cite{ebpf} monitor to observe system calls made by the application.

Although our prototype implementation targets Python, the underlying design is language-agnostic. Its core components rely on techniques that are generalizable to other ecosystems. For instance, the policy monitor could be adapted for PHP or JavaScript using different instrumentation backends, and the system call module is inherently cross-language.

\subsection{Parsing Annotations to Policy}
To integrate our annotations seamlessly into Python without altering the language's syntax, we implemented them to appear as standard function calls. The core of our implementation is an instrumentation hook on the \texttt{CALL\_FUNCTION} opcode, which handles function invocations.
When the \texttt{CALL\_FUNCTION} opcode is executed, our instrumentation code extracts the function name from the stack and checks if it is one of the annotations. If it is an annotation, the instrumented annotation parsing code will parse additional annotation options from the function arguments into the run-time policies. This is an implementation choice for convenience in Python, not a limitation of our overall design. The core principle of parsing developer-provided syntax to create security policies is language-agnostic.
The policy monitor then receives the parsed policy and enforces this policy during subsequent program execution.

\subsection{Vulnerability Detection}
Business logic vulnerabilities violate an application's intended semantic policies without causing program crashes so they cannot be detected by traditional sanitizers. 
We therefore propose a set of custom sanitizers using our annotation system, which enables the explicit definition and identification of these high-level policy violations to detect typical vulnerabilities in our scope.

\subsubsection{Sensitive Information Leakage}
Developers sometimes mistakenly output sensitive information into logs or exception statements, directly or indirectly. 
To detect such issues, users can use the \texttt{Data Flow} class of annotations to mark sensitive data sources (e.g., a variable holding a password) and data sinks (e.g., a logging function). 
Our system then taints the source variables and tracks the propagation of this taint to track the data flow of those sensitive variables.

The primary implementation challenge is to maintain taint propagation robustly across both pure Python code and native C/C++ extension modules. 
For the Python code, we modified the CPython interpreter by adding a taint attribute to the base PyObject struct. Since all Python objects inherit from PyObject, this allows any object to carry a taint mark. We then instrumented the CPython to check and propagate this taint attribute for each opcode, effectively tracking data flow during the Python execution~\cite{schwartz2010SoK}. 
For the cross-language interface, we instrument the argument-parsing functions used by C extension modules, ensuring that taint information is preserved when data moves from Python to C. 
For the code written in C/C++, we provide a modified version of the LLVM data flow sanitizer. Developers can apply this pass by compiling the extension modules written in C with additional compilation flags and LLVM passes provided by \tool to apply the modified data flow sanitizer.

\tool can also detect timing side-channel vulnerabilities by statistically testing whether functions that handle sensitive data run in constant time. A classic example in Python is using the ``\texttt{==}'' to compare the user-provided password against the stored password, the number of matching characters influences the execution time, allowing an attacker to discover the stored password.
A function is targeted for this analysis either automatically if it consumes a variable that has been marked sensitive via \texttt{Data Flow} and \texttt{Object Access} annotation or explicitly if a user uses \texttt{WATCH.CON} annotation.
\tool profile the execution time of functions across different inputs using the algorithm mentioned in Dudect~\cite{reparaz2017dude}, which is selected because of its precision and usability as analyzed by Fourné et al.~\cite {fourne2024these}. If the execution time varies with different inputs, \tool will report a potential timing side-channel vulnerability.

\subsubsection{Improper Access Control}

Improper access allows unauthorized users to get access to resources (e.g., admin page or sensitive data) that should only be accessible by privileged users. 
To detect improper access control vulnerabilities, the \texttt{Object Access} and \texttt{Code Execution} annotations can be applied to the source code. 

\texttt{Object Access} annotations entail adding the specified variable to a watch list, along with the privileges defined in the annotation. 
Instrumentation is then applied to value retrieval and storage opcodes (such as Load\_* and Store\_*) to monitor operations that involve access or modification of a particular variable. 
When a variable is added to the watch list, \tool initially assesses its scope (global or local). 
Upon execution of load/store opcodes, \tool compares the corresponding variable against those on the watch list.
This is done via the \texttt{oparg} in Python, which is utilized by the opcode to resolve to the actual argument. 
Notably, global variables are checked globally, while local variables are only examined within the stack frame where they are defined. 
Similarly, for Python dynamic modules written in C, \tool leverages \texttt{ptrace} to set read/write breakpoints to check whether the variable is changed at runtime. 
When a breakpoint is triggered, a handler function is invoked that sends a signal to the policy monitor system of \tool. 

In the scenario that the developer knows which part of the code should be accessible to the privileged users, the \texttt{Code Execution} annotation is provided to show that the code after the annotations should be privileged-user-only. 
\texttt{Code Execution} annotations are similar to a flipped assertion: If the program execution encounters this annotation and the expression in the argument evaluates to \texttt{true}, the program touches a code region that it should not execute. 

\subsubsection{Unintended System Calls}
As noted above, vulnerabilities like SSRF, path traversal, and unsafe deserialization all share the common trait of using system calls differently from the developer's intention. 
To detect these vulnerabilities, developers can use the \texttt{System Call} class of annotations to express the intended usages of system calls.  

To detect such vulnerabilities, we need to implement a system call monitor module. 
Using \texttt{LD\_preload} to hook the system call wrappers in \texttt{libc} is not comprehensive, as the modules compiled from C could directly invoke system calls. 
Therefore, we implement our tool \tool on top of \ac{ebpf} using the Aya~\cite{aya-rs} library in around 3,000 lines of Rust code. 
In detail, the system call monitor leverages \ac{ebpf} trace points which are a set of reference points or hooks that are attained as the kernel performs a certain task. 
The \ac{ebpf} program is attached to two events: \texttt{raw\_syscalls:sys\_enter} and \texttt{raw\_syscalls:sys\_exit}, representing the kernel is about to enter a system call and exit a system call, respectively. 
The system call ID and arguments are collected in the \texttt{raw\_syscalls:sys\_enter} event, and the return value is collected in the \texttt{raw\_syscalls:sys\_exit} event. 
Then, this collected system call metadata is added to a hash map.

When the interpreter runs into a \texttt{System Call} class of annotations, it parses the annotation to extract the policy. The extracted policy along with the interpreter's process ID is then sent to the system call policy monitor module.
The module attaches the \ac{ebpf} part to the system call trace point and begins to collect data.
Every time the hash map is updated, the system call monitor module compares the record against the policy to verify if there is a policy violation. If a policy violation is detected, a signal is sent back to the interpreter and the interpreter triggers a segmentation fault. 

\tool monitors all file-related system calls, and it is extensible to detect other types of vulnerabilities abusing system calls. One prime example is detecting file-based \ac{toctou} vulnerabilities.
One typical case of \ac{toctou} involves a file initially accessed by a system call that checks attributes like \texttt{access} or \texttt{stat}, followed by another system call that performs actions like writing or reading on the same file. Another type occurs when a program creates a file and, later, changes the privileges after operations, such as writing data to the file. 
Although this idiom causes a \ac{toctou} bug, it is still widely used. 
Thus, to focus on critical reports, we only detect \ac{toctou} on files containing sensitive information with the help of our data flow analysis.
We flag a potential \ac{toctou} vulnerability when either of these patterns occurs, along with data flow that involves reading or writing sensitive variables to or from the file. 

The ability to implement a new detector for \ac{toctou} by composing our system call monitoring with data flow policies demonstrates the extensibility of the \tool framework.

\section{Evaluation}
\label{sec:evaluation}

We evaluated our prototype implementation \tool in five different sets of experiments which show that \tool is capable of identifying various business-logic vulnerabilities listed in the CWE Top 40 in real-world applications, that its runtime overhead is minimal, and that the annotation system is easy to use even for first-time users. 
We use \tool together with the Python fuzzer Atheris~\cite{atheris} which we will refer to as \toolfuzz.
In this setup, during fuzzing with Atheris, \tool will promote policy violations to crashes which will be observable by the fuzzer.
All experiments were conducted on an Intel Core i9-13900K machine with 64 GB of RAM running Ubuntu 22.04.

In the following,
\begin{inparaenum}[(A)]
\item we assess the feasibility of rediscovering known, publicly disclosed business logic vulnerabilities. This study is conducted by an author proficient in both \tool and software security to demonstrate that a knowledgeable user can effectively annotate unfamiliar applications, and to confirm that \tool{}’s annotations are sufficiently expressive to cover a wide range of vulnerabilities.
\item We apply \toolfuzz to popular open-source Python-based applications to evaluate if the approach can find new 0-day business-logic vulnerabilities in popular and actively-maintained Python projects.
\item We compare \toolfuzz against the standard Atheris~\cite{atheris} fuzzer as an ablation study.
\item We perform a usability study to assess the annotation system. We begin with an annotation study recruiting undergraduate and graduate students with varying levels of security and software development expertise to assess the ease of writing annotations for detecting vulnerabilities in unfamiliar applications, establishing a lower bound on effectiveness. Complementing this, we assess the system's real-world applicability through a qualitative study with professional developers and security analysts, focusing on their current detection practices, their perceived utility of the tool, and the potential barriers to its adoption in industrial workflows.
\item We measure the runtime performance overhead of \tool.
\end{inparaenum}

The target applications used in this section include both web applications and standalone Python applications. For standalone Python applications, we use the test cases in the source repository of the respective applications and transform them into a harness for \toolfuzz. For web applications, we need to use a different strategy, as they receive network requests (instead of getting inputs directly in binary applications). Therefore, we implement a custom mutator for \toolfuzz to mutate fields in the network request. To make the generated test cases more likely to be valid, the mutator will focus on mutating cookies, query parameters, headers, and URLs. Hence, the fuzzer can generate test cases having an appropriate request format. For applications that need a valid login, \tool will keep the session cookie in each request to be able to test the endpoints after the login authentication. To accelerate the fuzzing process, the mutator has a dictionary of serialized malicious payloads and code injection payloads trying to invoke a shell that would be easily captured by the policy monitor mechanism.

\subsection{Rediscovering Known Vulnerabilities}
\label{sec:eval_rediscover}
This experiment assesses whether \tool can effectively detect various real-world business logic flaws when guided by \toolfuzz. We begin by curating a list of vulnerable Python packages from the Snyk database~\cite{snyk}, filtering for recent, high-impact vulnerabilities that fall within the CWE Top 40.
After excluding applications with unsupported dependencies or those that function solely as libraries (such as cryptographic libraries, see Appendix~\ref{sec:appendix:skip_app}), we established a final set of \numreproducedapplications diverse applications, as listed in Table~\ref{tab:table_APP}.
To avoid introducing bias by knowing the detailed vulnerability report, the author who conducts this evaluation only knows the bug type of the application, is only allowed to access the code of the application that should be annotated, and has access to the application's documentation.

\begin{table}
    \centering
    \sisetup{
        table-format=4.0,
        table-auto-round=true,
        drop-exponent=true,
        exponent-mode=fixed,
        fixed-exponent=3
    }
    \begin{threeparttable}
        \caption{Known Vulnerabilities in Rediscovery Experiment,
        With Annotation Type Used for Rediscovery and the Information Required for Annotation per Application}
        \label{tab:table_APP}
        \scriptsize

        \begin{tabular}{@{ }r@{ }l@{}r@{ }S[table-format=3.1]@{ \ }l@{ }S@{ }r@{ }c@{ \ }c@{}}
            \toprule

            {\bf ID} & {\bf Name}                                                                                    & {\bf CWE} & {\bf Stars} & {\bf Vuln. Identifier}          & {\bf LoC}  & {\bf Type}  & {\bf Info} \\
                     &                                                                                               &           & {$/ 10^3$}  &                                 & {$/ 10^3$} &             &            \\
            \midrule

            1        & \href{https://github.com/jjjake/internetarchive}{internetarchive}                             & 362       & 1500        & \textsc{SNYK-6141253}           & 7782       & DF          & Docs       \\

            2        & \href{https://github.com/Backblaze/b2-sdk-python}{b2-sdk-python}                              & 362       & 166         & \textsc{CVE-2022-23651}         & 35116      & DF          & Docs       \\

            3        & \href{https://github.com/Backblaze/B2_Command_Line_Tool}{B2\_Command\_Line\_Tool}             & 362       & 522         & \textsc{CVE-2022-23653}         & 15302      & DF          & Docs       \\

            4        & \href{https://github.com/langchain-ai/langchain}{langchain}                                   & 918       & 80100       & \textsc{CVE-2024-2057}          & 362740     & SC          & Const      \\

            5        & \href{https://github.com/langchain-ai/langchain}{langchain}                                   & 918       & 80100       & \textsc{CVE-2024-0243}          & 362740     & SC          & Const      \\

            6        & \href{https://github.com/HumanSignal/label-studio}{label-studio}                              & 918       & 16100       & \textsc{CVE-2023-47116}         & 32641      & SC          & Const      \\

            7        & \href{https://github.com/benbusby/whoogle-search}{whoogle-search}                             & 918       & 8700        & \textsc{CVE-2024-22205}         & 3298       & SC          & Const      \\

            8        & \href{https://github.com/gradio-app/gradio}{gradio}                                           & 918       & 27800       & \textsc{SNYK-6141123}           & 63982      & SC          & Const      \\

            9        & \href{https://github.com/PaddlePaddle/Paddle}{Paddle}                                         & 22        & 21500       & \textsc{CVE-2024-0818}          & 1055529    & SC          & Const      \\

            10       & \href{https://github.com/daswer123/xtts-api-server}{xtts-api-server}                          & 22        & 179         & \textsc{SNYK-6398416}           & 2606       & SC          & Const      \\

            11       & \href{https://github.com/langchain-ai/langchain}{langchain}                                   & 22        & 80100       & \textsc{CVE-2024-28088}         & 362740     & SC          & Const      \\

            12       & \href{https://github.com/esphome/esphome}{esphome}                                            & 22        & 7400        & \textsc{CVE-2024-27081}         & 371356     & SC          & Const      \\

            12       & \href{https://github.com/onnx/onnx}{onnx}                                                     & 22        & 16600       & \textsc{CVE-2024-27318}         & 120972     & SC          & Const      \\

            13       & \href{https://github.com/HumanSignal/label-studio}{label-studio}                              & 434       & 16100       & \textsc{SNYK-6347239}           & 32641      & SC          & Const      \\

            14       & \href{https://github.com/zenml-io/zenml}{zenml}                                               & 434       & 3600        & \textsc{CVE-2024-28424        } & 187468     & SC          & Const      \\

            15       & \href{https://github.com/inventree/InvenTree}{inventree}                                      & 434       & 3600        & \textsc{CVE-2022-2111}          & 99204      & SC          & Const      \\

            16       & \href{https://github.com/StanfordVL/GibsonEnv}{GibsonEnv}                                     & 502       & 817         & \textsc{CVE-2024-0959}          & 31212      & SC          & Docs       \\

            17       & \href{https://github.com/huggingface/transformers}{Transformers}                              & 502       & 123000      & \textsc{CVE-2023-6730}          & 1154503    & SC          & Docs       \\

            18       & \href{https://github.com/vanderschaarlab/synthcity}{synthcity}                                & 502       & 341         & \textsc{CVE-2024-0936}          & 46310      & SC          & Docs       \\

            19       & \href{https://github.com/apache/airflow}{Apache-Airflow}                                      & 862       & 34000       & \textsc{CVE-2023-50944}         & 658410     & OA          & Docs       \\

            20       & \href{https://github.com/dgtlmoon/changedetection.io}{changedetection.io}                     & 306       & 14600       & \textsc{CVE-2024-23329}         & 15860      & CE          & Docs       \\

            21       & \href{https://github.com/hyperledger/aries-cloudagent-python}{aries-cloudagent}               & 287       & 385         & \textsc{CVE-2024-21669}         & 213203     & OA          & Docs       \\

            22       & \href{https://github.com/MobSF/Mobile-Security-Framework-MobSF}{MobSF}                        & 276       & 16100       & \textsc{CVE-2023-42261}         & 23480      & OA          & Docs       \\

            23       & \href{https://github.com/mlflow/mlflow}{mlflow}                                               & 287       & 17100       & \textsc{CVE-2023-6014}          & 239221     & OA          & Docs       \\

            24       & \href{https://github.com/janeczku/calibre-web}{calibre-web}                                   & 863       & 11300       & \textsc{CVE-2022-0405}          & 29774      & CE          & Docs       \\

            25       & \href{https://github.com/wagtail/wagtail}{wagtail}                                            & 200       & 17100       & \textsc{CVE-2023-45809}         & 186030     & DF          & API        \\

            26       & \href{https://github.com/ansible/ansible}{ansible-core}                                       & 20        & 60700       & \textsc{CVE-2024-0690}          & 203432     & DF          & API        \\

            27       & \href{https://github.com/pyload/pyload}{pyLoad}                                               & 200       & 3100        & \textsc{CVE-2024-21644}         & 52103      & DF          & API        \\

            28       & \href{https://github.com/omise/omise-python}{omise}                                           & 200       & 26          & \textsc{SNYK-6138437}           & 7971       & DF          & API        \\

            29       & \href{https://pypi.org/project/apache-airflow-providers-celery/}{airflow-providers-celery}    & 200       & 34000       & \textsc{CVE-2023-46215}         & 658410     & DF          & API        \\

            30       & \href{https://github.com/openstack/horizon}{horizon}                                          & 601       & 1300        & \textsc{CVE-2020-29565}         & 113963     & SC          & Const      \\

            31       & \href{https://github.com/evennia/evennia}{evennia}                                            & 601       & 1700        & \textsc{SNYK-6591326}           & 150592     & SC          & Const      \\

            32       & \href{https://github.com/pyload/pyload}{pyLoad}                                               & 601       & 3100        & \textsc{CVE-2024-24808}         & 52103      & SC          & Const      \\

            33       & \href{https://github.com/run-llama/llama_index}{llama\_index}                                 & 94        & 31700       & \textsc{CVE-2024-3098}          & 305184     & SC          & Docs       \\

            34       & \href{https://github.com/aimhubio/aim}{Aim}                                                   & 94        & 4800        & \textsc{CVE-2024-2195}          & 27552      & SC          & Docs       \\

            35       & \href{https://github.com/vantage6/vantage6}{vantage6}                                         & 94        & 24          & \textsc{CVE-2024-21649}         & 34752      & SC          & Docs       \\

            36       & \href{https://github.com/DIRACGrid/DIRAC}{DIRAC}                                              & 668       & 108         & \textsc{CVE-2024-29905}         & 225072     & SC          & Docs       \\

            37       & \href{https://github.com/fonttools/fonttools}{fonttools}                                      & 611       & 4100        & \textsc{CVE-2023-45139}         & 171624     & SC          & Docs       \\

            38       & \href{https://github.com/geopython/OWSLib}{OWSLib}                                            & 611       & 369         & \textsc{CVE-2023-27476}         & 29003      & SC          & Docs       \\

            39       & \href{https://github.com/stchris/untangle}{untangle}                                          & 611       & 607         & \textsc{CVE-2022-31471}         & 704        & SC          & Docs       \\

            40       & \href{https://github.com/zopefoundation/AccessControl}{AccessControl}                         & 269       & 12          & \textsc{CVE-2024-51734}         & 8621       & CE          & Docs       \\

            41       & \href{https://github.com/jupyter/jupyter_core}{Jupyter\_Core}                                 & 427       & 197         & \textsc{CVE-2022-39286}         & 3451       & SC          & Docs       \\

            42 & \href{https://github.com/clearml/clearml}{clearml} 
                             & 522         & 6.2        & \textsc{CVE-2024-24595} &  187
                             & DF & Docs \\

            43 & \href{https://github.com/indico/indico/}{indico}
                            &  639            & 1.9 & \textsc{CVE-2024-50633} & 117 
                            & OA & Docs \\

            44       & \href{https://github.com/vantage6/vantage6}{vantage6}                                         & 287       & 24          & \textsc{CVE-2024-21653}         & 34752      & OA\tnote{*} & ---        \\

            45       & \href{https://github.com/apache/superset}{Apache-Superset}                                    & 287       & 57800       & \textsc{CVE-2023-27526}         & 175207     & OA\tnote{*} & ---        \\

            46       & \href{https://github.com/zenml-io/zenml}{zenml}                                               & 287       & 3600        & \textsc{CVE-2024-25723}         & 187468     & OA\tnote{*} & ---        \\

            47       & \href{https://github.com/nautobot/nautobot-app-device-onboarding}{nautobot-device-onboarding} & 200       & 35          & \textsc{CVE-2023-48700}         & 3091       & DF\tnote{*} & ---        \\
            
            \bottomrule
        \end{tabular}
        \begin{tablenotes}
            \item[] SC: System Call, DF: Data Flow, OA: Object Access, CE: Code Execution
            \item[*] \textbf{Not rediscovered} due to lack of information
        \end{tablenotes}
    \end{threeparttable}

\end{table}

To evaluate an application with \tool, we first create a docker container for the application, allowing for easier reproducibility. 
If we are testing access control vulnerabilities, the fuzzer is only given a low-privilege role, such as a guest user (without registration) instead of a logged-in user or a default user instead of an administrator account. 
Otherwise, we provide the fuzzer with a privileged account to test the application more comprehensively. 
Next, we add the required annotations. 
This procedure took the authors of this work a total of ten working days.
Finally, we fuzz each application with \tool for 24 hours.
If we fail to detect the vulnerability, we check the vulnerability report to analyze the cause of our failure. 
Table~\ref{tab:table_APP} lists metadata for each application along with the vulnerability identifier and the required annotation.
The column \emph{Info} indicates the knowledge used to place the annotation correctly: \emph{Const} indicates knowledge of a constant variable, e.g., confining file accesses to a root directory defined as a constant.
\emph{Docs} indicates understanding the context of the application via the documentation.
For example, applications with authenticated endpoints should only be accessible to authorized users.
Finally, \emph{API} indicates knowledge of third-party API documentation, e.g., that the data returned by an API is sensitive.
The asterisk indicates vulnerabilities that could not be reproduced without further knowledge, as described below.

\toolfuzz can successfully rediscover \numreproducedbugs vulnerabilities.
Notably, in the case of \texttt{label-studio} (ID 6), not only was the vulnerability detected, but using the same annotation, a bypass for the official patch was also discovered.
The patch relies on the remote user to determine the file type, allowing an obfuscated extension name to bypass the check.

\toolfuzz was unable to detect four vulnerabilities (IDs 44--47 in Table~\ref{tab:table_APP}).
The first is in \texttt{vantage6} (ID 44), which has an insecure default SSH configuration that allows root access using only password authentication.
While password-less authentication is recommended, we did not consider this to be the flagged security issue and, thus, did not add an annotation for it. Otherwise, \toolfuzz could detect this issue. 
Similarly, \texttt{apache-superset} (ID 45) has a fine-grained privilege control system, which we did not have enough understanding of to create a meaningful annotation. 
In \texttt{zenml} (ID 46), the vulnerability was not triggered due to complex preconditions, where user A needs to create an invitation link and send it to user B. 
Then, the invited user B will register and craft a request to update user A's password. 
We verified that \toolfuzz could theoretically detect this vulnerability by manually providing a seed containing a short valid invitation link, showing the importance of the input corpus. 
In \texttt{nautobot-device-onboaring} (ID 47), we successfully added the annotations to the correct place, but the vulnerability will only be triggered under a particular configuration, with which \toolfuzz also finds the vulnerability.

\result{
We successfully reproduce most (43 / 47) of the vulnerabilities only with knowledge of the bug type and project documentation.
In the cases where we missed vulnerabilities, maintainers would likely have a higher success rate due to their deeper understanding of the project's business logic.
}

\subsection{Finding New Vulnerabilities with \toolfuzz}
\label{sec:eval_zeroday}
To evaluate \toolfuzz's ability to find previously unknown vulnerabilities in real-world applications, we conducted a study on real-world open-source applications collected from GitHub.
We selected a set of 60 popular and actively maintained Python projects from GitHub (over 200 stars and recently committed). To test our approach against hardened targets, this set included four projects with active bug bounty programs (Apache Superset, Apache Airflow, ZenML, Calibre-Web) and one that undergoes regular security audits (Home Assistant Core)~\cite{homeassistantSecurityAudits}.

For each project, we simulated a user's workflow with a one-hour time limit to understand the application's core logic and write corresponding annotations. If a project could not be understood within this timeframe, it was skipped. We then fuzzed each annotated application using a harness crafted from its existing test cases. 
All findings were manually verified.

We reported a total of \numzerodays issues. Of these, \numvlunids were assigned vulnerability identifiers. 17 vulnerabilities are assigned CVE IDs and the remaining 3 confirmed issues received internal tracking identifiers from Apache, GitHub, and Microsoft, respectively (one each).
Table~\ref{tab:table_APP_0day} provides a detailed list of these vulnerabilities\footnote{Assigned identifiers are removed to ensure an anonymous submission.}.
Since an issue is usually found by one individual annotation, we indicate its \emph{Type}.
To quantify the required human effort, the table contains the \emph{Time} it took us to create the annotations and the number of each type of annotation we created.

As the table shows, the human effort required was modest; the time to analyze the code and write the necessary annotations was approximately 60 minutes per application, on average. 
This demonstrates that a user with prior knowledge of \tool can become effective on a new and unfamiliar codebase very quickly. 
In Section~\ref{sec:eval_usability}, we discuss more about the usability of \tool when the user has no prior knowledge of both \tool and the target applications.
\begin{table}[tb]
    \centering
    \sisetup{
      table-format=4.0,
      table-auto-round=true,
      drop-exponent=true,
      exponent-mode=fixed,
      fixed-exponent=3
    }
    \begin{threeparttable}
      \caption{New Zero-day Vulnerabilities Found by \tool}
      \label{tab:table_APP_0day}
      \scriptsize
      \begin{tabular}{@{ }r@{ \ }l@{}r@{ \ }S[table-format=2.1]@{ \ }S[table-format=4]@{ \ }c@{ }r*{4}{r@{ \ }}@{ }}
        \toprule
        {\bf ID} & 
        {\bf Name} & 
        {\bf CWE} & 
        {\bf Stars} & %
        {{\bf LoC}} & 
        {\bf Type} & 
        {\bf Time} & 
        \multicolumn{4}{c}{\bf \# Annotations} \\
        \cmidrule(lr){8-11}
        &  &  & {$/ 10^3$} & {$/ 10^3$} &  & $/ \text{min}$ & {SC} & {DF} & {OA} & {CE} \\
        \midrule

        1 & \href{https://github.com/xlang-ai/OpenAgents}{OpenAgents} & 434& 3300 & 12730 &SC& 60  & 12 & 8 & 11 & 16\\ 

        2 & \href{https://github.com/spiral-project/ihatemoney}{ihatemoney} & 732 & 1100 & 9073 & OA & 30  & 7 & 3 & 10 & 12 \\ 

        3 & \href{https://github.com/home-assistant/core}{Home Assistant core} & 532& 68000 & 1629234 & DF & 70  & 14 & 9 & 18 & 25 \\ 

        4 & \href{https://github.com/apache/airflow}{Apache-Airflow} & 367& 33900 & 658410 & DF & 90  & 19 & 8 & 14 & 19 \\ 

        5 & \href{https://github.com/vastsa/FileCodeBox}{FileCodeBox} & 532& 3000 & 4952 & DF & 45  & 8 & 4 & 6 & 12 \\ 

        6 & \href{https://github.com/nebari-dev/nebari}{nebari} & 532& 251 & 14387 & DF & 50  & 6 & 5 & 3 & 12 \\ 

        7 & \href{https://github.com/CloudOrc/SolidUI}{SolidUI} & 532& 494 & 4568 & DF & 75  & 13 & 7 & 2 & 13 \\ 

        8 & \href{https://github.com/WordOps/WordOps}{WordOps} & 532& 1200 & 14050 & DF & 40  & 3 & 8 & 2 & 2 \\ 

        9 & \href{https://github.com/WordOps/WordOps}{WordOps} & 367& 1200 & 14050 & DF & 40  & 3 & 8 & 2 & 2 \\ 

        10 & \href{https://github.com/ArchiveBox/ArchiveBox}{ArchiveBox} & 367& 19300 & 12968 & DF & 20  & 6 & 2 & 0 & 0\\ 

        11 & \href{https://github.com/apache/superset}{Apache-Superset} & 434& 57500 & 175207 & SC & 80  & 16 & 4 & 10 & 15 \\ 

        12 & \href{https://github.com/veops/cmdb}{cmdb} & 434& 1200 & 19851 & SC & 40  & 9 & 6 & 8 & 13 \\ 

        13 & \href{https://github.com/zenml-io/zenml}{zenml} & 367& 3600 & 189796 & DF & 65  & 11 & 4 & 7 & 15 \\

        14 & \href{https://github.com/cpacker/MemGPT}{MemGPT} & 208& 8600 & 25339 & DF & 40  & 10 & 6 & 2 & 3 \\ 

        15 & \href{https://github.com/binux/pyspider}{pyspider} & 208& 16300 & 14761 & DF & 55  & 4 & 5 & 8 & 6 \\ 

        16 & \href{https://github.com/alandtse/alexa\_media\_player}{alexa\_media\_player} & 532& 1300 & 6757 & DF & 30  & 2 & 3 & 1 & 0 \\

        17 & \href{https://github.com/Fannovel16/comfyui\_controlnet\_aux}{comfyui\_controlnet\_aux} & 94& 1400 & 195727 & SC & 70  & 10 & 5 & 0 & 2\\

        18 & \href{https://github.com/lithops-cloud/lithops}{lithops} & 94& 306 & 28915 & SC & 45  & 13 & 3 & 13 & 3 \\

        19 & \href{https://github.com/Kedreamix/Linly-Talker}{Linly-Talker} & 94& 676 & 36521 & SC & 50  & 6 & 7 & 0 & 0 \\

        20 & \href{https://github.com/cms-sw/cmssw}{cmssw} & 94 & 1100 & 1596828 & SC & 95  & 20 & 7 & 0 & 2 \\

        21 & \href{https://github.com/microsoft/RecAI}{Microsoft RecAI} & 94 & 368 & 28939 & SC & 45  & 5 & 3 & 0 & 0 \\

        22 & \href{https://github.com/janeczku/calibre-web}{calibre-web} & 434 & 11300 & 29774 & SC & 85  & 22 & 7 & 13 & 8 \\

        \bottomrule
      \end{tabular}

      \begin{tablenotes}
        \item[] SC: System Call, DF: Data Flow, OA: Object Access, CE: Code Execution
      \end{tablenotes}
    \end{threeparttable}
\end{table}

\result{
Empirically, we found that with moderate human effort, \tool can discover 22 previously unknown vulnerabilities (17 CVE assigned), even in hardened targets with active bug bounty programs.
}

\textbf{False Positives.}
We found two false-positive sensitive information leakage vulnerabilities during this experiment.
The first false positive is in the \href{https://github.com/codemation/easyauth}{\texttt{EasyAuth}} project.
We added a data flow annotation for a variable \texttt{token} containing a sensitive authentication token.
Indeed, this token is written to a log file, which is detected by \toolfuzz.
However, the token is revoked before logging, making this a false positive.
This issue could have been avoided if we had added the token revocation as a taint-removing function. 
The second false positive occurred in the \href{https://github.com/munki/munki}{\texttt{Munki}} project, which writes the authorization token, to a \href{https://github.com/curl/curl}{\texttt{curl}} configuration file created by \texttt{mkstemp()}. 
We confirmed with the developer that this is the intended behavior and the temporary file is destroyed after use. 
Both cases are due to our limited understanding of the projects and should be no hindrance for the developers.
 
\subsection{Ablation Study}
\label{sec:eval_compare}
To evaluate \tool's capabilities, we conducted experiments on a benchmark suite of 35 business logic vulnerabilities (e.g., path traversal, access control flaws) curated from four well-established, intentionally vulnerable Python applications: OWASP's Pygoat~\cite{pygoat}, OWASP Vulnerable Flask App~\cite{owaspvulnflask}, Damn Small Vulnerable Web (DSVW)~\cite{DSVW}, and The Vulnerable API (VAmPI)~\cite{vampi}. This suite, comprising $\approx19,200$ lines of code, provides a ground truth for measuring detection accuracy of our approach.

To evaluate \tool's effectiveness as a sanitizer, we conducted an ablation study that isolates its core contribution.
We aim to demonstrate that \tool provides a standard fuzzer with the necessary sanitizer to detect business logic vulnerabilities that it would otherwise miss.
To this end, we compared \toolfuzz (Atheris + \tool) against the baseline fuzzer (Atheris without an additional sanitizer) on our benchmark suite.
The results clearly show that our approach significantly enhances the fuzzer’s ability to uncover logic flaws:
\toolfuzz successfully detected all 35 vulnerabilities.
In contrast, the baseline fuzzer detected none, failing even when directly supplied with the exact inputs known to trigger the flaws.
This result demonstrates that a standard fuzzer is blind to these vulnerability classes and that \tool provides the essential oracle capability required for their detection.

Next, we also tried to benchmark \toolfuzz against other established tools to evaluate its performance. A direct comparison with other sanitizers was not feasible, as no tools with \tool's similar capability exist. Other fuzzing tools were also unsuitable for a fair comparison due to being overly specialized for single bug types (e.g., EDEFuzz~\cite{edefuzz}, CT-Fuzz~\cite{he2020ct} SSRFuzz~\cite{SSRFuzz}), non-available preconditions (e.g., EDEFuzz~\cite{edefuzz} for GUI, FLOWFUZZ~\cite{infoflowfuzz} for perfect resetting or relying on non-portable, heuristic-based methods for different language ecosystems (e.g., Atropos~\cite{Atropos} for PHP, ODDFuzz~\cite{ODDFuzz} for Java).

Therefore, we compared \toolfuzz against the most relevant state-of-the-art static and dynamic vulnerability scanners. In summary,  \toolfuzz significantly outperforms all scanners in finding the 35 target vulnerabilities. A detailed breakdown of each scanner's performance and evaluation is available in Appendix~\ref{sec:appendix:scanner_details}.

\result{In summary, \tool provides an essential and previously missing sanitizer to standard fuzzers and can also empower fuzzers to outperform existing state-of-the-art scanners by integrating direct developer insight into a dynamic state monitoring framework.
}

\subsection{User Studies}
\label{sec:eval_user_study}

\subsubsection{Annotation Study}
\label{sec:eval_usability}
\begin{table*}
    \centering
    \caption{Usability Study Result}
    \label{tab:user_study}

    \scriptsize
    \begin{tabular}{c *{6}{r@{ }r@{ }r@{ \ }c}}
        \toprule
        & \multicolumn{4}{c}{\bf \href{https://github.com/gradio-app/gradio}{Gradio}} & \multicolumn{4}{c}{\bf \href{https://github.com/daswer123/xtts-api-server}{xtts-api-server}} & \multicolumn{4}{c}{\bf \href{https://github.com/veops/cmdb}{cmdb}} & \multicolumn{4}{c}{\bf \href{https://github.com/vanderschaarlab/temporai}{temporai}} & \multicolumn{4}{c}{\bf \href{https://github.com/WordOps/WordOps}{WordOps}} & \multicolumn{4}{c}{\bf \href{https://github.com/dgtlmoon/changedetection.io}{changedetection.io}} \\
        & \multicolumn{4}{c}{SSRF} & \multicolumn{4}{c}{Path Traversal} & \multicolumn{4}{c}{Unrestricted Upload} & \multicolumn{4}{c}{Untrusted Deserial.} & \multicolumn{4}{c}{Information Leakage} & \multicolumn{4}{c}{Broken Access Control} \\ 
        \cmidrule(lr){2-5} \cmidrule(lr){6-9} \cmidrule(lr){10-13} \cmidrule(lr){14-17} \cmidrule(lr){18-21} \cmidrule(lr){22-25}
        \textbf{ID} & Time & Num & (FP) & Result & Time & Num & (FP) & Result & Time & Num & (FP) & Result & Time & Num & (FP) & Result & Time & Num & (FP) & Result & Time & Num & (FP) & Result \\
        & $/ \text{min}$ & & & & $/ \text{min}$ & & & & $/ \text{min}$ & & & & $/ \text{min}$ & & & & $/ \text{min}$ & & & & $/ \text{min}$ & & & \\
        \midrule
        P1 & 58 & 20 & 1 & \checked & 21 & 6 & 0 & \checked & 49 & 3 & 0 & \checked & 22 & 5 & 0 & \checked & 20 & 17 & 1 & \checked & 82 & 23 & 1 & \xmark \\
        P2 & 52 & 10 & 2 & \checked & 22 & 8 & 0 & \checked & 55 & 12 & 0 & \checked & 49 & 5 & 0 & \checked & 45 & 19 & 1 & \checked & 89 & 19 & 0 & \xmark \\
        P3 & 90 & 76 & 4 & \checked & 10 & 10 & 1 & \checked & 60 & 46 & 3 & \checked & 20 & 3 & 0 & \checked & 60 & 40 & 1 & \checked & 90 & 68 & 9 & \checked \\
        P4 & 120 & 16 & 0 & \checked & 30 & 10 & 0 & \checked & 90 & 10 & 0 & \checked & 20 & 5 & 0 & \checked & 75 & 16 & 1 & \checked & 60 & 8 & 0 & \checked \\
        P5 & 140 & 69 & 2 & \checked & 43 & 41 & 5 & \checked & 66 & 127 & 18 & \checked & 35 & 15 & 0 & \checked & 93 & 90 & 7 & \checked & 45 & 64 & 7 & \xmark \\
        P6 & 50 & 36 & 0 & \checked & 30 & 13 & 1 & \checked & 35 & 72 & 0 & \checked & 20 & 3 & 0 & \checked & 65 & 62 & 3 & \checked & 40 & 16 & 2 & \checked \\
        P7 & 180 & 36 & 3 & \checked & 80 & 15 & 1 & \checked & 80 & 76 & 0 & \checked & 38 & 5 & 1 & \checked & 70 & 38 & 1 & \checked & 180 & 93 & 3 & \checked \\
        P8 & 60 & 54 & 0 & \checked & 40 & 20 & 2 & \checked & 60 & 123 & 11 & \checked & 40 & 24 & 3 & \checked & 35 & 62 & 0 & \checked & 60 & 69 & 0 & \xmark \\
        P9 & 120 & 33 & 2 & \checked & 30 & 3 & 0 & \xmark & 120 & 61 & 3 & \checked & 90 & 11 & 0 & \checked & 90 & 21 & 0 & \checked & 180 & 69 & 1 & \xmark \\
        P10 & 60 & 13 & 0 & \checked & 30 & 21 & 0 & \checked & 30 & 23 & 1 & \xmark & 30 & 1 & 0 & \checked & 30 & 14 & 0 & \xmark & 60 & 13 & 1 & \xmark \\
        P11 & 60 & 22 & 1 & \xmark & 40 & 13 & 0 & \checked & 40 & 17 & 1 & \checked & 60 & 2 & 0 & \checked & 150 & 29 & 1 & \checked & 70 & 5 & 0 & \xmark \\
        \cmidrule(lr){2-5} \cmidrule(lr){6-9} \cmidrule(lr){10-13} \cmidrule(lr){14-17} \cmidrule(lr){18-21} \cmidrule(lr){22-25}
        \textbf{Avg.} & 86 & 33 & & & 40 & 15 & & & 62 & 46 & & & 45 & 11 & & & 63 & 28 & & & 80 & 41 & & \\
        \bottomrule
    \end{tabular}
\end{table*}

To evaluate the usability and effectiveness of the annotations in \tool, we conducted a user study with 11 voluntarily recruited participants from two institutions located in different countries. 
These participants had no prior experience with \tool or the target applications.
To mitigate the large time investment required for human subject experiments, we used a subset of the applications randomly selected from Table~\ref{tab:table_APP} and Table~\ref{tab:table_APP_0day} covering unique vulnerability types, rather than the full dataset. 
Each participant was tasked with annotating these six applications with which they were unfamiliar.
This scenario was designed to be more challenging than a developer annotating their own codebase.
We measured the time taken, the number of annotations, and the effectiveness of the participants at detecting the known vulnerability.

As detailed in Table~\ref{tab:user_study}, the participants demonstrated a high degree of effectiveness despite their lack of familiarity with the target applications. Across 66 total tasks (11 participants × 6 applications), the user-provided annotations successfully detected the target vulnerabilities in 55 cases, yielding an 83.3\% success rate.
We observed a very low false positive rate, indicating that the annotations lead to precise policies. Key feedback from participants highlighted that \tool's annotation syntax was straightforward to learn and that they found it more practical to use blocklists to forbid specific bad behaviors rather than creating comprehensive allowlists. The most common failure point (7 of 11) were on a single, unusually difficult access control vulnerability in \texttt{changedetection.io}. Getting this annotation right is difficult, as this flaw required deep, non-local knowledge of the application's intended logic regarding API authorization. 
Still, four participants’ annotations successfully detected this bug.
Note that this kind of vulnerability is not an ideal target for \tool.
This is because if the developer is aware that this endpoint should be protected, they can directly fix the issue without the need for testing.
Other failures stemmed from overlooking documentation (e.g., \texttt{WordOps}) or misunderstanding a specific vulnerability (e.g., SSRF in \texttt{Gradio}).

The results in Table~\ref{tab:user_study} and the self-reported knowledge levels indicate that users with basic conceptual knowledge of development were able to use \tool to write annotations for detecting vulnerabilities in a new application. Additional knowledge in application development or security vulnerabilities further enhances the annotation quality. We believe that if the real user who is the developer uses \tool, the time would be shorter and the detection rate would be higher with fewer false positives. A detailed breakdown of the study methodology, participant demographics, failure cases, false-positive case analysis, and feedback is provided in Appendix~\ref{sec:appendix:user_study_details}.

\subsubsection{Real-world Developer Study}
To assess the real-world applicability and adoption potential of \tool, we conducted a qualitative study with 10 software developers and security analysts, representing a wide range of professional experience. The full methodology, participant demographics, and questionnaire are detailed in Appendix~\ref{sec:appendix:real_world_study_details} and Table~\ref{tab:questionnaire}.

The study has three key findings.
First, it confirmed the problem's relevance: all (10/10) participants had encountered business-logic vulnerabilities, and their detection process relies heavily on manual code review (10/10) and external penetration testing (9/10), with only one participant (1/10) having used static analysis for this purpose.
This validates that the problem remains a significant, work-intensive challenge.
Second, we found universal agreement on \tool's value proposition.
When asked about the effort-versus-value trade-off, participants unanimously (10/10) agreed that spending approximately one hour adding annotations to a project was a reasonable exchange for the ability to detect business-logic vulnerabilities in applications with critical functionalities.
Third, the primary barriers to adoption identified by participants were of practical nature: the learning curve of a new annotation language (10/10) and the effort of integration with CI/CD pipelines (6/10).
This feedback validates the perceived value of \tool's approach and provides clear, practical directions for future work centered on tooling, usability, and workflow integration.

\result{
Our user studies collectively demonstrate both the practical usability and real-world value of \tool. The annotation study demonstrates that effective annotations can be created with minimal prior knowledge of \tool and the target application. The real-world study with professional developers showed that the annotation effort required for our approach is considered a reasonable trade-off for its security value.}

\subsection{Performance Evaluation}
\label{sec:eval_performance}
Finally, we evaluate the performance overhead of \tool. 
The runtime performance is measured by using the Python Performance Benchmark Suite~\cite{Pythonperfbench}, which is an authoritative benchmark for alternative Python implementations provided. 
This fits our purpose, as \tool is implemented by modifying the CPython interpreter. 
All benchmarks are executed in rigorous mode to get accurate data. 
The taint tracking module is evaluated by tainting every variable, an extreme setup that is not typical for regular usage. 
Additionally, we measure the syscall module overhead by recording every system call. Details can be found in Appendix~\ref{sec:appendix:performance_eval}.

\result{
We find that tracing all variables results in an average runtime overhead of 10\%, while tracing all system calls leads to an overhead of about 5\%. 
}

\section{Discussion and Future Work}
\label{sec:discuss}

Our prototype implementation of \tool enables developers to express their intuition of the program behavior with annotations.
In this section, we differentiate \tool from binary analysis frameworks and explore potential future extensions regarding automated annotations and language support.

Our prototype implements a practical engineering approach to cross-language taint analysis, which proved effective in our evaluation.
We acknowledge that comprehensive cross-language taint analysis is a challenging research problem.
A more robust implementation could build upon existing work, such as PolyCruise~\cite{polycruise} for the Python-C interface and other works addressing different language pairs~\cite{crosslanguagetaintjava, crosslanguagetaintmulti}.

\begin{table}
  \centering

  \begin{threeparttable}

    \caption{Comparative Analysis of \tool, Valgrind, and DynamoRIO}
    \label{tab:dbi_comparison}
    \scriptsize
    \rowcolors{2}{white}{gray!10}
    \newcolumntype{Y}{>{\centering\arraybackslash}X}

    \begin{tabularx}{\linewidth}{p{4.2em}YYY}

      \toprule
      \textbf{Tool} & \textbf{\tool} & \textbf{Valgrind} & \textbf{DynamoRIO} \\
      \midrule

      \textbf{Goal} & Human-in-the-loop sanitizer framework 
       & \multicolumn{2}{>{\hsize=\dimexpr2\hsize+2\tabcolsep}Y}{\acf*{dbi} framework for building binary analysis tools}\\

      \textbf{Target} & Source code & \multicolumn{2}{c}{Binary} \\

      \textbf{Interface} & Policies defined via intuitive annotations in source code 
      &  \multicolumn{2}{>{\hsize=\dimexpr2\hsize+2\tabcolsep}Y}{C API on IR / instructions for building instrumentation tools to manipulate execution} \\

      \textbf{User} &  App. developer / Security analyst & \multicolumn{2}{c}{Security expert}  \\

      \textbf{Vuln.} & Business logic vulnerabilities & 
      \multicolumn{2}{>{\hsize=\dimexpr2\hsize+2\tabcolsep}Y}{Enables detection of any machine-level behavior; pre-built tools cover low-level execution errors (memory safety violations)} \\

      \textbf{Overhead} & \emph{Low:} 10\%/5\% & \emph{High:} 5,417\%\tnote{*} & \emph{High:}  5,513\%\tnote{*} \\ 
      \bottomrule

    \end{tabularx}
    \begin{tablenotes}
      \item [*] Lower bound of the performance overhead measured for memory tracing on the Python Performance Benchmark Suite as detailed in Appendix~\ref{sec:appendix:dbi:dbi-perf}
    \end{tablenotes}
  \end{threeparttable}

\end{table}

\subsection{Comparison with \acs*{dbi} Frameworks}
As \tool is at its core a framework to build custom sanitizers, it might seem similar to \ac{dbi} frameworks like Valgrind~\cite{valgrind} and DynamoRIO~\cite{dynamorio} that provide an API to build custom low-level binary analysis tools.
However, they represent a \emph{fundamentally} different design paradigm, %
as summarized in Table~\ref{tab:dbi_comparison}.
There is a strong divergence in the semantic information available, the target users, and the performance characteristics (a more detailed comparison is provided in Appendix~\ref{sec:appendix:dbi}).
\subsubsection{The Semantic Gap}
\ac{dbi} frameworks operate at the machine code layer, observing instructions rather than application objects like variables. 
They view the execution of a Python program as a stream of instructions from the interpreter binary, remaining blind to the high-level logic those instructions represent.
For example, a \ac{dbi} tool cannot directly identify a \texttt{password} object in memory.
To implement \tool's functionality, a \ac{dbi} user would need to reverse engineer the CPython interpreter's internal memory layout during runtime.
This is an enormous challenge, as structural information is lost in translation (cf. the challenges in implementing binary sanitizers~\cite{binarysan}).
In contrast, \tool operates directly within the CPython interpreter, granting it access to the semantic context from an internal perspective.
\subsubsection{Target Users}
As \tool is designed for application developers and security analysts with  knowledge of the target application, users can add intuitive annotations to detect vulnerabilities (as shown in Section~\ref{sec:eval_user_study}) without understanding the internal details of \tool.
In contrast, \ac{dbi} frameworks require users to also possess expert knowledge of binary analysis, systems programming and the framework's internal APIs to build custom analysis tools.
\subsubsection{Performance}
While it would be theoretically possible to implement \tool's functionality using \ac{dbi} frameworks, their low-level instrumentation comes with high runtime overhead.
Memory tracing alone, the first step to implementing something like \tool via \ac{dbi}, introduces a prohibitive overhead of over 5,000\% (50x), which is orders of magnitude higher than \tool{}'s overhead in our evaluation (as detailed in Appendix~\ref{sec:appendix:dbi:dbi-perf}).

\medskip
In conclusion, \ac{dbi} frameworks are ill-suited for detecting business logic vulnerabilities, particularly in high-level languages like Python.
\tool provides the necessary abstraction and efficiency for sanitizing business logic vulnerabilities, especially in the performance-sensitive context of fuzzing.

\subsection{Integrating \acp{llm}}
Annotations require manual effort to analyze the application and the quality of these annotations depends on the user's expertise, which can be a challenge in large codebases or when dealing with unfamiliar libraries.
A potential solution is integrating \acp{llm} for semi-automating the generation of these policies.
An \ac{llm} could reduce manual effort by analyzing source code to propose relevant security annotations and bridge the expertise gap for complex dependencies.
However, utilizing \acp{llm} is not a simple solution due to challenges like hallucinations and the required large context windows to process sufficient documentation.
These limitations necessitate that any \ac{llm}-generated annotations undergo rigorous validation by a human expert to ensure their trustworthiness~\cite{10.1145/3728894}.

Our work establishes the foundational framework to express these policies.
We view this framework as the necessary first step to transform the semantic information from developer knowledge into machine-enforceable policies, enabling and supporting future research into reliable automation with \acp{llm}.

\subsection{Porting to Other Languages}
While \tool can be extended to other languages, such as PHP and JavaScript, to find business logic vulnerabilities, we view this primarily as an engineering challenge, as the fundamental strategy of \tool is language-agnostic.
We briefly sketch the porting strategy for each specific component of the framework:
The \textit{Annotation System} syntactically resembling standard function calls
could be ported by adding built-in functions or instrumenting the language-specific function invocation mechanism to intercept these calls. Similar to our approach for Python's \texttt{CALL\_FUNCTION} opcode, a PHP implementation could target \texttt{DO\_FCALL} or \texttt{DO\_ICALL}, and JavaScript could target the \texttt{Call} opcode.
The \textit{System Call Monitor} is implemented using eBPF for Linux kernel hooks, agnostic to the language runtime and therefore directly reusable without modification.
The \textit{Data Flow Monitor} could leverage existing dynamic taint tracking solutions such as PHP's Taint extension, Augur~\cite{aldrich2022augur}, taintflow~\cite{taintflow} for JavaScript, or libDFT~\cite{libdft} for C/C++.
The \textit{Object Access Monitor} requires instrumenting low-level operations for variable access, like PHP's \texttt{ZEND\_FETCH\_R/W/RW} opcodes, or the property load (\texttt{Lda*}) and store (\texttt{Sta*}) instruction families in JS engines.

\section{Related Work}
\label{sec:related_work}
Security vulnerabilities detection is a vast research area. In this section, we focus on the most relevant prior work in dynamic analysis and code sanitization, particularly concerning non-memory safety and business logic flaws.

Dynamic Sanitizers and System Policies. Established dynamic sanitizers like AddressSanitizer~\cite{serebryany_addresssanitizer_2012}, MemorySanitizer~\cite{stepanov_memorysanitizer_2015}, LeakSanitizer~\cite{llvmLeakSanitizerx2014}, UBSan~\cite{undefinedbehaviorsanitizer_2013} and ThreadSanitizer~\cite{serebryany2009threadsanitizer} are highly effective for memory corruption and data races but do not target business logic vulnerabilities. System-level policy enforcement mechanisms like  Landlock~\cite{landlockLandlockUnprivileged} or AppArmor~\cite{apparmorAppArmor} are also distinct, as they enforce coarse-grained, process-wide policies, lacking the fine-grained, runtime-adjustable control needed for specific code blocks or contexts relevant to business logic. 
Recent sanitizers for non-memory flaws are often limited, targeting narrow vulnerability types (e.g., numerical errors~\cite{10.1145/3446804.3446848}, code injections~\cite{patil2016automated}), issues unique to embedded systems~\cite{liu2024effectively} or restricted to specific languages like Go~\cite{wang2019go} or PHP~\cite{Atropos}. We also differentiate \tool from tools detecting application-specific correctness bugs~\cite{10.1145/3485533, liang2022detecting, artzi2010finding, eshghieHighGuardCrossChainBusiness2024}. (e.g., broken HTML~\cite{artzi2010finding} or misinterpreted SQL~\cite{liang2022detecting}), which find functional errors rather than the security-critical policy violations \tool targets.

Many fuzzing frameworks integrate custom bug oracles, but these are often limited to specific vulnerability classes (e.g., XSS~\cite{webfuzz, witcher}, injection vulnerabilities~\cite{witcher}, SSRF~\cite{SSRFuzz}) or languages (e.g., Atropos for PHP~\cite{Atropos}). Tools targeting information leaks also have different limitations. EDEFuzz~\cite{edefuzz} can only detects sensitive data exposure in API responses but lacks the business-logic context to determine if data is truly sensitive or check leakage from other channels like log files. Ct-fuzz~\cite{he2020ct} focuses on low-level side-channel leaks, distinct from \tool's whole-system logic analysis. FLOWFUZZ~\cite{infoflowfuzz} requires complex, deterministic setups and manual instrumentation for data leak detection, and does not cover timing side-channels. Other tools target narrow issues like file uploads~\cite{uchecker, fuse, huang2021ufuzzer} or domain-specific policies like robotics~\cite{kim2021pgfuzz}. In short, while prior fuzzing frameworks target specific, predefined bug patterns, \tool provides a general framework for defining and detecting violations of application-specific policies.

Finally, several approaches leverage developer input or annotations. IJON~\cite{9152719} uses annotations to guide fuzzers towards deeper application states but still relies on traditional bug oracles (e.g., crashes) rather than detecting new vulnerability classes. ASIDE~\cite{7357200} and Anovul~\cite{ghorbanzadeh2020anovul} use annotations or code markers to check for access control and authentication flaws, respectively. This concept is similar to \tool's \texttt{Code Execution} annotation. However, \tool distinguishes itself through a significantly broader and more generalizable annotation framework designed to specify and detect a wide spectrum of complex business logic vulnerabilities, extending far beyond access control or state reachability goals.

In summary, while previous work has covered various aspects of business-logic vulnerability detection, \tool introduces a novel, annotation-driven dynamic sanitization approach specifically designed to identify security-critical business logic flaws, filling a crucial gap in existing defenses.

\section{Conclusion}
\label{sec:conclusion}
In this paper, we present a novel, annotation-based sanitization framework to address the critical challenge of detecting business logic vulnerabilities. 
Based on our analysis of existing fuzzing sanitizers, we find that current sanitizers often rely on brittle, automated heuristics that cannot capture the necessary application-specific semantic context. To overcome this, \tool empowers developers to directly express an application's intended security policies using a lightweight and intuitive annotation system. 
By encoding a developer's implicit knowledge into explicit, machine-readable annotations, we open up new classes of vulnerabilities for (semi-)automated bug discovery via dynamic code analysis.
To this end, we propose a set of annotations that cover the business logic vulnerabilities in CWE's Top 40 most dangerous software weaknesses.

Our prototype, integrated with a standard fuzzer, called \toolfuzz, demonstrates the effectiveness of this approach by rediscovering \numreproducedbugs known and detecting \numzerodays previously unknown vulnerabilities in popular, well-maintained open-source projects.
A total of \numcves CVE identifiers are assigned to our findings at the time of writing. Our annotation study and performance benchmarks further confirmed that the system is easy to use and incurs minimal overhead.

By shifting the focus from inferring behavior with brittle heuristics to enforcing explicitly defined policies, \tool represents a step forward in the detection of business logic vulnerabilities. We believe this paradigm of leveraging direct developer insight provides a powerful and extensible foundation for securing the complex applications of today and tomorrow.
We hope that our research helps push fuzzing, a proven effective bug-finding technique for low-level programming, into the realm of high-level programming languages.

\section{Ethics Considerations}
For the annotation study and real-world developer study, we followed best practices regarding surveys and studies. Before we conducted the user study, our study plan and related materials were evaluated and approved by the ethical review board (ERB) of the authors' affiliated institution.
The full instruction document will be published as part of our research artifact. During the study, the participants could voluntarily opt out of the study at any point in time. A document describing the purpose of the study was presented to the participants before they agreed or declined to join. After data collection, we deleted all the data that could identify individual participants.

For the new vulnerabilities discovered by \toolfuzz, we follow best practices and ethical standards for handling the vulnerabilities identified during our evaluation. Following a \emph{coordinated disclosure} approach, we have reported all \numzerodays vulnerabilities discovered by \tool to the developers using the protocols set out in their security advisories. We are actively working with the developers to help them understand the vulnerabilities and submit pull request proposals to fix them.

\section{Acknowledgment}
The project underlying this paper was funded with funds from the Federal Ministry of Transport (BMV) under the funding code 45AVF5A011. The author is responsible for the content of this publication.

\printbibliography

\begin{appendices}
\label{sec:appendix}

\section{Target Vulnerabilities}
\label{sec:appendix:target_vuln}
\begin{table*}[!htbp]
  \scriptsize
  \centering
  \caption[]{\tool's support for the CWE Top 40 Security Weaknesses. ``\faCheck'' shows supported vulnerability classes, while ``\faTools'' refers to vulnerabilities where an annotation system is not useful, as they can be directly patched if the developer has the correct intuition. Furthermore, ``\faBook'' shows classes supported by complimentary tools. This existing work is listed in the last column, even though these tools mostly focus on technical, non-business logic-related issues. Business Logic Needed (BLN) refers to whether the bugs can only be found with ($\CIRCLE$) or strictly without ($\Circle$) an understanding of the application's business logic. There are vulnerability classes where only some bugs require this understanding ($\LEFTcircle$).}
  \label{tab:table_cwe} 
\begin{tabularx}{\linewidth}{r l l l c c X}

  \toprule
  {\bf Rank} & {\bf CWE} & {\bf Description} & {\bf Exploitation} & {\bf BLN} & {\bf \tool} & {\bf Existing Work}  \\
  \midrule

\rowcolor{white} 1  & 79  & Cross-site Scripting & APP Data& $\Circle$ & \faBook & FuzzOrigin~\cite{281314fuzzorigin}, KameleonFuzz~\cite{KameleonFuzz}, webFuzz~\cite{webfuzz}  \\
\rowcolor{white} 2  & 787 & Out-of-bounds Write & Memory& $\Circle$ & \faBook & ASan~\cite{serebryany_addresssanitizer_2012} \\
\rowcolor{white} 3  & 89  & SQL Injection & DataBase &$\Circle$  & \faBook & Witcher~\cite{witcher}, Atropos~\cite{Atropos}  \\
\rowcolor{white} 4  & 352 & Cross-Site Request Forgery (CSRF)  & System Call&$\LEFTcircle$ & \faTools & WebFuzzAuto~\cite{csrffuzz} \\
\rowcolor{gray!25} 5  & 22 & Path Traversal  & System Call& $\CIRCLE$ & ~\faCheck~ & Atropos~\cite{Atropos}, PHUZZ~\cite{PHUZZ} \\
\rowcolor{white} 6  & 125 & Out-of-bounds Read  & Memory & $\Circle$ & \faBook & ASan~\cite{serebryany_addresssanitizer_2012} \\
\rowcolor{white} 7  & 78 & OS Command Injection & System Call & $\LEFTcircle$ & ~\faCheck~ & Witcher~\cite{witcher}, Atropos~\cite{Atropos}  \\
\rowcolor{white} 8  & 416 & Use After Free  & Memory & $\Circle$ & \faBook & ASan~\cite{serebryany_addresssanitizer_2012} \\
\rowcolor{tableblue} 9 & 862 & Missing Authorization &Object Access/Code Exec. & $\CIRCLE$ & ~\faCheck~ & ~ \\
\rowcolor{gray!25} 10 & 434 & Unrestricted Upload of File & System Call&  $\LEFTcircle$ & ~\faCheck~ & Atropos~\cite{Atropos}, UFuzzer~\cite{ufuzzer}, URadar~\cite{URadar}\\
\rowcolor{tableblue} 11 & 94 & Code Injection &System Call/Data Flow & $\LEFTcircle$ & ~\faCheck~ & ~ \\
\rowcolor{white} 12  & 20 & Improper Input Validation & System Call & $\LEFTcircle$ & \faTools & Witcher~\cite{witcher}, Atropos~\cite{Atropos} \\
\rowcolor{white} 13 & 77 & Command Injection &System Call & $\LEFTcircle$ & ~\faCheck~ & Witcher~\cite{witcher}, Atropos~\cite{Atropos}  \\
\rowcolor{tableblue} 14 & 287 & Improper Authentication  &Object Access/Code Exec. & $\CIRCLE$ & ~\faCheck~ & ~ \\
\rowcolor{tableblue} 15 & 269 & Improper Privilege Management&Object Access/Code Exec. & $\CIRCLE$ & ~\faCheck~ & ~ \\
\rowcolor{gray!25} 16 & 502 & Deserialization of Untrusted Data  &System Call & $\CIRCLE$ & ~\faCheck~ & ODDFuzz~\cite{ODDFuzz}, PHUZZ~\cite{PHUZZ} \\
\rowcolor{gray!25} 17 & 200 & Exposure of Sensitive Information  &Data Flow/System Call & $\CIRCLE$ & ~\faCheck~ & EDEFuzz~\cite{edefuzz}, FLOWFUZZ~\cite{infoflowfuzz} \\
\rowcolor{tableblue} 18 & 863 & Incorrect Authorization  &Object Access/Code Exec. & $\CIRCLE$ & ~\faCheck~ & ~ \\
\rowcolor{white} 19 & 918 & Server-Side Request Forgery (SSRF) &System Call & $\LEFTcircle$ & ~\faCheck~ & Atropos~\cite{Atropos}, SSRFuzz~\cite{SSRFuzz}\\
\rowcolor{white} 20 & 119 & Improper Restriction of Ops.\ w/i Mem.\ Buffer  & Memory &  $\Circle$ & \faBook & ASan~\cite{serebryany_addresssanitizer_2012} \\
\rowcolor{white} 21 & 476 & NULL Pointer Dereference & Memory &  $\Circle$ & \faBook & ASan~\cite{serebryany_addresssanitizer_2012} \\
\rowcolor{white} 22 & 798 & Use of Hard-coded Credentials& Data & $\CIRCLE$ & \faTools & ~ \\
\rowcolor{white} 23 & 190 & Integer Overflow or Wraparound  & Memory &  $\Circle$ & \faBook & UBSan~\cite{undefinedbehaviorsanitizer_2013}  \\
\rowcolor{white} 24 & 400 & Uncontrolled Resource Consumption  & DOS & $\Circle$ & ~\faCheck~ & All Fuzzers \\
\rowcolor{tableblue} 25 & 306 & Missing Authentication for Critical Function  &Object Access/Code Exec. & $\CIRCLE$ & ~\faCheck~ & ~ \\
\rowcolor{white} 26 & 770 & Allocation of Resources Without Limit  & DOS & $\Circle$ & \faBook & All Fuzzers \\
\rowcolor{gray!25} 27 & 668 & Exposure of Resource to Wrong Sphere &Object Access/Code Exec. & $\CIRCLE$ & ~\faCheck~ & EDEFuzz~\cite{edefuzz} \\ 
\rowcolor{white} 28 & 74 & Improper Neutralization of Special Elements & System Call/Data Flow & $\LEFTcircle$ & ~\faCheck~  & Witcher~\cite{witcher} \\
\rowcolor{tableblue} 29 & 427 & Uncontrolled Search Path Element & System Call & $\LEFTcircle$ & ~\faCheck~ & ~ \\
\rowcolor{tableblue} 30 & 639 & Authorization Bypass  &Object Access/Code Exec. & $\CIRCLE$ & ~\faCheck~ & ~ \\
\rowcolor{gray!25} 31 & 532 & Insertion of Sensitive Information into Log File & Data Flow & $\CIRCLE$ & ~\faCheck~ & FLOWFUZZ~\cite{infoflowfuzz} \\
\rowcolor{tableblue} 32 & 732 & Incorrect Permission Assignment &Object Access/Code Exec. & $\CIRCLE$ & ~\faCheck~ & ~ \\
\rowcolor{white} 33 & 601 & Open Redirect  &System Call & $\LEFTcircle$ & ~\faCheck~ & OpenRedireX~\cite{OpenRedireX} \\
\rowcolor{white} 34 & 362 & Race Condition & System Call/Object Access & $\LEFTcircle$ & ~\faCheck~ & TSan~\cite{serebryany2009threadsanitizer}, CONZZER~\cite{CONZZER}, krace~\cite{krace} \\
\rowcolor{tableblue} 35 & 522 & Insufficiently Protected Credentials &Data Flow/Object Access & $\CIRCLE$ & ~\faCheck~ & ~ \\
\rowcolor{white} 36 & 276 & Incorrect Default Permissions&Object Access/Code Exec. & $\CIRCLE$ & \faTools & ~ \\
\rowcolor{gray!25} 37 & 203 & Observable Discrepancy & Data Flow & $\CIRCLE$ & ~\faCheck~ & CT-Fuzz~\cite{he2020ct} \\
\rowcolor{white} 38 & 59 & Link Following &System Call & $\Circle$& \faTools & ~ \\
\rowcolor{white} 39 & 843 & Type Confusion & Memory & $\Circle$  & \faBook & type-san~\cite{type-san} \\
\rowcolor{white} 40 & 312 & Cleartext Storage of Sensitive Information & Data Flow & $\Circle$  & \faBook & ~ \\

  \bottomrule

  \end{tabularx} 

\end{table*}

We list the Top 40 Most Dangerous Software Vulnerabilities from the Common Weakness Enumeration (CWE) project~\cite{mitreMostDangerous} in Table~\ref{tab:table_cwe}. 
We also list previous sanitizers and similar generalizable approaches for fuzzing in the table.
Note that we only mention typical tools for each already supported bug class, as, for example, for memory corruption issues a multitude of tools exists~\cite{sanitization-sok}. 
Traditionally, fuzzers have been able to detect memory corruption-related and other technical vulnerabilities (marked with $\Circle$) by leveraging known sanitizers like \ac{asan}~\cite{serebryany_addresssanitizer_2012} and \ac{ubsan}~\cite{undefinedbehaviorsanitizer_2013}.
Witcher~\cite{witcher} targets injection-related vulnerabilities like command injection and FuzzOrigin~\cite{281314fuzzorigin} targets cross-site scripting for web applications.
EDEFuzz~\cite{edefuzz} detects excessive data exposures on API endpoints. 

\tool can work together with these existing fuzzing technologies to extend the types of vulnerabilities (marked with \cmark), that can be detected via fuzzing. 
There are a few unsupported vulnerabilities (marked with $\otimes$) that could be directly patched if the developer has the correct intuition (e.g., for SQL statements, a developer can turn the SQL query into a prepared statement to mitigate such vulnerabilities). 
Thus, annotations for these cases would be redundant. 
An example is CWE-798 (Use of Hard-coded Credentials): If the developer knows that this is a bad practice, they could directly patch the code without the need of any fuzzing or other testing. 
Similarly, \tool cannot help detect Cross-Site Request Forgery (CSRF) because if the developer is aware of it, a CSRF token can prevent the issue. 
Finally, CWE-770 (Allocation of Resources Without Limits or Throttling) is neither supported by \tool nor other existing work, as most instances of this bug class result in timeouts, that are detected by most fuzzers by default. 
However, in some cases annotations could benefit the detection of this bug class, as we discussed in Section~\ref{sec:discuss} of this work.

\section{Comparison with State-of-the-art Scanners}
\label{sec:appendix:scanner_details}
\begin{table}[!htbp]
  \centering

  \begin{threeparttable}
    \caption{Comparative Evaluation Against SOTA Tools }
    \label{tab:tool_cmp}

    \scriptsize
    \setlength{\tabcolsep}{3pt}
    \begin{tabular}{@{ }l *{4}{r@{ }r} r @{ \; }r@{ }}
      \toprule
      & \multicolumn{2}{c}{\bf Pygoat} & \multicolumn{2}{c}{\bf FLASK} & \multicolumn{2}{c}{\bf DSVW} & \multicolumn{2}{c}{\bf VAmPI} & {\bf Precision} & {\bf Recall} \\
      & \multicolumn{2}{c}{$P = 16$} & \multicolumn{2}{c}{$P = 5$} & \multicolumn{2}{c}{$P = 10$} & \multicolumn{2}{c}{$P = 4$} \\
      \cmidrule(lr){2-3} \cmidrule(lr){4-5} \cmidrule(lr){6-7} \cmidrule(lr){8-9}
                 & TP & FP & TP & FP & TP & FP & TP & FP &         &         \\
      \midrule
      Semgrep    &  7 &  2 &  3 &  1 &  4 &  0 &  1 &  0 &  83.3\% &  42.9\% \\
      SonarQube  & 10 &  0 &  2 &  0 &  6 &  0 &  0 &  0 & 100.0\% &  51.4\% \\
      Pysa       &  6 &  1 &  3 &  0 &  2 &  0 &  1 &  0 &  92.3\% &  34.3\% \\
      ZAP        &  5 &  1 &  4 &  0 &  6 &  0 &  2 &  1 &  94.1\% &  48.6\% \\
      Wapiti     &  3 &  9 &  2 &  0 &  3 &  1 &  0 &  2 &  40.0\% &  22.9\% \\
    Atheris    &  0 &  0 &  0 &  0 &  0 &  0 &  0 &  0 &  0\% & 0\% \\
      \addlinespace
    {\bf \toolfuzz}& 16 &  0 &  5 &  0 & 10 &  0 &  4 &  0 & 100.0\% & 100.0\% \\
      \bottomrule
    \end{tabular}
    \begin{tablenotes}
      \item[] TP: True Positive, FP: False Positive, P: Positive
      \item[] Precision = TP / (TP + FP), Recall = TP / P
    \end{tablenotes}

  \end{threeparttable}
\end{table}

We compare \toolfuzz against three static and two dynamic analysis tools that share at least three supported vulnerability types with \toolfuzz and are actively maintained.
Note that the selected projects are well-established tools used for comparison in other academic works~\cite{Atropos, witcher, lee2022link, sastUsed01}.
We compare against the static tools Semgrep~\cite{semgrep}, SonarQube~\cite{sonarqube}, and Meta's Pysa~\cite{pysa}, as well as the dynamic tools Zed Attack Proxy (ZAP)~\cite{zap} and Wapiti~\cite{wapiti}.

All tools were configured according to their official documentation, which for tools like Pysa and ZAP involved significant manual effort to define sources, sinks, and access rules. These static tools were provided with the same data source and sink information as our annotations. Additionally, we used existing Semgrep rules. We did not perform extensive manual configuration (e.g., project-wide type annotations for Pysa), as this would by far exceed the typical manual effort required for \tool and is infeasible for us.

Table~\ref{tab:tool_cmp} contains the results of this experiment, including the number of true and false positive reports for each tool, as well as precision and recall.
We find that \toolfuzz reports all 35 bugs in scope within the 24-hour fuzzing trial, while the second-place SonarQube reports 18 bugs. 
Only SonarQube and \toolfuzz produce no false positive reports.
The best dynamic web vulnerability scanner, ZAP, performs similarly to SonarQube, with 17 reports. %
The results highlight the fundamental limitations of existing tools and showcase the core strength of \tool: combining dynamic analysis with human-provided semantic context.

As already observed by Güler et al.~\cite{Atropos}, our findings also suggest that dynamic web scanners are limited in their bug-finding capabilities. The dynamic scanners (ZAP, Wapiti) struggled significantly. As black-box tools, they analyze HTTP responses with heuristics and often fail to achieve deep code coverage. \tool's policy monitor has visibility into the application's internal state and logic during execution. For instance, ZAP's access control checks are not granular enough to distinguish between read and write permissions, causing it to miss several vulnerabilities that \tool detected easily. While evaluating ZAP, we noticed that correctly configuring the vulnerability detection rules requires in-depth knowledge of ZAP. 
We expect an annotation-based approach to be more intuitive for developers.

The static analysis tools (Semgrep, Pysa) produced a high number of false negatives, failing to detect many vulnerabilities. Their lack of runtime information forces them to approximate program behavior, leading to imprecision and false positives. Regarding Pysa, as its functionality focuses on static taint analysis, the analysis needs to be approximated, which causes some false positive reports. Pysa's documentation highlights that it demands both source/sink configuration files and extensive manual code annotations for accuracy. Manual annotation helps prevent issues like ``taint collapsing''---where tracking many attributes or keys leads to excessive tainting---but fully annotating projects was impractical. Our approach involved using a thorough configuration file while only annotating code sections pertinent to the vulnerabilities. Pysa also struggles with opaque code (like C extensions or bytecode), defaulting to assuming taint propagation from arguments to return values. \tool only needs the source/sink information and overcomes this limitation by gathering accurate information via dynamic code execution.

\section{Comparison with \ac{dbi} Frameworks}
\label{sec:appendix:dbi}

\tool is fundamentally different from general-purpose \ac{dbi} frameworks like Valgrind~\cite{valgrind} and DynamoRIO~\cite{dynamorio}. While all are frameworks developed for dynamic analysis, they are not competing technologies. As summarized in Table~\ref{tab:dbi_comparison}, they are complementary tools with starkly different designs, target users, and objectives.

\subsection{Fundamental Design Difference}
The core design of \tool is that of a sanitizer framework. Its purpose is to provide an intuitive and user-friendly system for application developers to define and enforce application-specific security policies directly in the source code. 

In contrast, Valgrind and DynamoRIO are machine-code instrumentation platforms. They are designed for tool-builders and systems programmers. Their purpose is to provide a comprehensive, low-level API to monitor and modify a program's binary instruction stream at runtime, enabling the creation of a wide range of custom binary analysis tools (e.g., profilers, memory checkers) that operate on binary executables.

\subsection{Interfaces and Ease of Use}
The intended user of each framework dictates its interface. For \tool, the user is the application developer or security analyst. The interface consists of high-level, function-like annotations embedded directly in the source code. The task is policy definition, not systems programming to implement program analysis tools. The user does not need to know the internal details of \tool.

For \ac{dbi} frameworks, the user is a tool-builder who must implement analysis logic using complex APIs provided by the frameworks. This interface provides powerful, low-level abstractions, such as Valgrind's ISA-independent Intermediate Representation (IR) or DynamoRIO's direct machine instruction stream manipulation, but requires deep expertise in the framework's internal architecture. Then the user can leverage such powerful APIs to further develop analysis tools.

\subsection{Vulnerability Detection Capabilities}
The frameworks themselves do not detect vulnerabilities as they provide the mechanisms that enable the vulnerability detection. \tool as a sanitizer specifically designed for business-logic vulnerabilities, does not target low-level errors supported by pre-built tools shipped with Valgrind/DynamoRIO. The \ac{dbi} frameworks, by design, are more general, can theoretically detect business-logic vulnerabilities but requiring additional expertise and engineering effort from the user to implement specialized tools.

For \tool, as a sanitizer framework, enforces the semantic policies defined by its annotation interface. It is therefore designed to detect violations of business-logic vulnerabilities. Valgrind, however, has a different goal, it's core capability is lifting binary code to an IR and providing support for shadow values. This design enables the creation of tools that detect violations of universal execution rules like memory safety (e.g., MemCheck) or threading rules (e.g., Helgrind). Similarly, DynamoRIO's core capability is providing a fast, fine-grained, and robust API for any manipulation of the machine instruction stream. 
This general-purpose design enables a vast range of tools which are mostly designed to find low-level execution errors like Dr.Memory for memory related errors. The rich APIs provided make it possible to further implement binary analysis tools to detect business-logic vulnerabilities.

\subsection{Limitations of Implementing \tool Using \ac{dbi} Frameworks}
While one might theoretically attempt to implement \tool's functionality on a \ac{dbi} framework, it is practically infeasible due to the semantic gap. To support a language like Python, \ac{dbi} frameworks operate on the compiled, machine-code binary of the CPython interpreter, not the high-level application logic it is executing. They are blind to the application's semantic context. 

For example, consider \tool's annotation \texttt{EXECUTION.BLOCK (user.type!='admin')}.
Via inernal interpreter instrumentation, \tool has direct native access to the user object and its type attribute. It operates at the application's semantic level. A \ac{dbi} tool, in contrast, would see only the \texttt{mov}, \texttt{cmp}, and \texttt{je} machine instructions of the CPython interpreter's binary. It has no direct concept of a user object. To acquire this information, the \ac{dbi} tool would have to monitor all memory accesses and attempt to reverse-engineer the CPython interpreter's memory layout to map raw addresses to Python run-time objects. This approach is inefficient and unreliable.

Similarly, a \ac{dbi}-based taint tracker is semantically blind. It can track data flow from memory address A to B, but it lacks the developer-provided intent to know that the variable \texttt{pwd} at address A is sensitive and the function print at address B is a dangerous sink in this specific context. Furthermore, known incompatibilities between CPython's custom memory allocator and tools like Memcheck can introduce a high rate of false positives~\cite{valgrindpymalloc}.

\subsection{Performance Overhead and Design Tradeoffs}
\label{sec:appendix:dbi:dbi-perf}
The semantic gap is not the only barrier. The performance overhead of a \ac{dbi}-based approach would be prohibitive. 
Same as the benchmark suited used in Section~\ref{sec:eval_performance}, we use The Python Performance Benchmark Suite~\cite{Pythonperfbench} which is designed to be an authoritative source of benchmarks for all Python implementations and focuses on real-world benchmarks with 87 different tasks in total. 
We evaluated the cost of monitoring all memory accesses, a necessary prerequisite for the reverse-engineering to map the memory to runtime objects. 
This monitoring introduces a runtime overhead of 5417\% (54.17x) with DynamoRIO's memory trace, 5513\% (55.13x) with Valgrind's Memcheck. Detailed results are in Table~\ref{tab:dbi_perf}. 
Additionally, we assess a memory tracing tool based on a slimmed down version of Valgrind's Lackey tool~\cite{valgrindlackey} with an overhead of 8756\% (87.56x). 
This exceeds Memcheck's overhead, as Lackey prioritizes implementation clarity over performance~\cite{valgrindlackey}. These results suggest that DBI-based implementations will have high overhead and developing them requires substantial expertise regarding the framework's internals.

This overhead is a direct result of different design tradeoffs. \tool trades generality (having source / interpreter access) for performance and semantic awareness.
Our current prototype is built on top of standard Python interpreters for the projects with source code available, the instrumented code can directly monitor and manipulate the Python run-time events from the internal perspective without need of inefficient reverse-engineering.
Its lightweight, targeted mechanisms are activated only when and where a specific event is triggered (e.g., annotation enable), resulting in a overhead of about 10\%. 
\ac{dbi} frameworks trade performance for capability and generality. Their heavyweight, JIT-based architecture is extremely expensive while they can run on any binaries. 
Moreover, they serve as a base framework providing rich APIs to the tool-builders, those tool-builders could future extend the functionalities by developing new tools using their knowledge on the \ac{dbi} frameworks and target problems.

While one could apply \tool's concepts using \ac{dbi} frameworks, the performance penalty makes it impractical for tasks like fuzzing, which rely on repeated program execution. 
A more suitable use case for a \ac{dbi}-based approach would be in single-run analyses, such as verifying a specific exploit on a binary-only target, where performance is not the primary concern.

\section{Annotation Study Details}
\label{sec:appendix:user_study_details}

A total of 15 participants initially expressed interest in joining the study. However, 4 were excluded because they were first- or second-year undergraduate students with no reported knowledge of application development or security analysis. This background was deemed too distant from \tool's target users, who are application developers and security analysts.
The 11 participants, referred to as P1--P11, were computer science undergraduate and graduate students with varying levels of expertise in application development and security.
Among them, P1--P2 had hands-on experience with security vulnerabilities and a basic understanding of application development.
P3--P5 were experts in security vulnerabilities with basic knowledge of application development.
P6--P8 had practical experience with web applications and a basic understanding of security vulnerabilities.
Lastly, P9--P11 possessed only a basic conceptual understanding of application development, with no prior knowledge of security vulnerabilities.
All of them do not have any prior experience with \tool or familiarity with the target code bases used in the study.
Each participant was tasked with annotating six randomly selected vulnerable applications. 
For each participant, we recorded the number of annotations added and the time it took to complete the task. Furthermore, we analyzed the effectiveness of the added annotations.
Moreover, we collected knowledge-level information and feedback from the participants.

\subsection{Preparation}
As the participants are unfamiliar with the applications, contrary to a developer who would use \tool, the participants are given the following information for the study: a summary of the application's main purpose as initial guidance, instructions on using \tool including the syntax of the potential annotations, and example annotations for each type of vulnerability. 
Furthermore, we provide information on common code patterns associated with each vulnerability type to help them understand these classes of vulnerabilities. 
To help them better understand where to add annotations, we instructed participants can begin with identifying security-critical boundaries where data crosses trust domains. 
We advised them that effective annotations are often derived from sources like documentation or API specifications to verify if a behavior is intended. 
Note that we avoided giving details about specific APIs or internal application knowledge from the applications used in the user study to prevent leaking hints to the participants. 
No information about the existing vulnerabilities is given to the participants. 
The full instruction document will be published as part of our research artifact in the anonymous repository.

\subsection{Participant Feedback}
Participants were asked to provide feedback after completing their assignments and to self-report their knowledge levels. The key takeaways from their feedback are summarized as follows: Concerns About Overlooking Critical Annotations: Participants expressed apprehension about potentially missing important locations to annotate, which led them to spend additional time during the initial phase of the process. Preference for Block Lists Over Allow Lists: Participants found it more comfortable and practical to use block lists rather than allow lists, as block lists were easier to construct and reduced the likelihood of false positives. Ease of Learning and Using Annotation Syntax: All participants reported that the syntax of \tool's annotations was straightforward to learn and use. 

Nine participants expressed concerns about potentially overlooking critical locations to annotate, leading them to spend considerable time minimizing this risk during the initial phase.
This concern is understandable, as the participants lacked detailed knowledge of the applications under test.
However, this issue is unlikely to arise for developers, who are familiar with the code bases they typically work on, even for large projects.

Ten participants' feedback mentions they were more confident while writing blocklists for system calls compared to writing allowlists.
This outcome was expected, as it is impractical for participants to have comprehensive knowledge of the underlying code.
Using blocklists to constrain code behavior proved to be both easier and effective, enabling participants to identify most vulnerabilities in the applications as shown in Table~\ref{tab:user_study}.
Differently, it is also mentioned that when the code or documentation shows explicit information like constant variables representing that a certain file path or an executable binary could be accessed, participants prefer to use the allowlist since it is more accurate.

All participants found annotation syntax of \tool straightforward to learn and use.
They described an initial learning curve, followed by a systematic approach to annotation.
Participants adopted a strategy of focusing on one annotation type at a time to reduce context switching.
For example, a participant might first identify and annotate sensitive variables (e.g., secrets), then move on to annotate potential system call sites, proceeding sequentially in this manner.
\subsection{Failure Cases}
P10 failed to write an annotation to detect the sensitive information leakage vulnerability in \texttt{WordOps} (10 of the 11 participants succeeded in this task), a WordPress stack management tool set. P10 missed the descriptions of credentials in the WordPress documentation and thus failed to detect this vulnerability. 
P9 failed to write an annotation to detect the path traversal in \texttt{xtts-api-server} (10 of the 11 participants succeeded in this task) because P9 restricted the wrong file path while using the file access annotation in the correct place.
P11 failed to write an annotation to detect the SSRF in \texttt{Gradio} (10 of the 11 participants succeeded in this task) because the participant had a misunderstanding of the SSRF vulnerability.
Regarding \texttt{cmdb}, P5 and P8 added more annotations compared to the other two participants.
This is because P5 and P8 add annotations at the beginning of each function protected by the authentication decorator, while the other participants add the annotation inside the decoration function. From a semantic point of view, both approaches are equivalent.

\subsection{False Positives}
To study the false positive cases, we manually inspect the annotations and show the data in the column FP in Table~\ref{tab:user_study}.
The number of false positives is low, and we find that the following two factors might help reduce the number even further:
the participants tend to write annotations around the code that explicitly express the developer's intention.
For example, annotations are added based on the conditions in the if-statement or a path concatenation which shows the folder the code will operate on.
The participants prefer to use blocklists (over allowlists), which is less likely to cause false positives. 

False positives in \tool are commonly associated with system call and data flow annotations.
System call-related false positives often occur within functions that contain a few lines of annotated code, whereas data flow-related false positives may appear in modules different from those where the annotations are made.
In either case, these false positives are straightforward for \tool's target users—application developers—to understand and resolve.
Developers are familiar with the codebase and can address these issues iteratively, refining annotations as needed to eliminate false positives effectively.
Furthermore, since our participants did not have access to the testing results, they could not fix false positives before submitting their annotations.
In reality, adding annotations is an iterative approach, and developers can quickly remove them.

\section{Feedback From Real-world Developers}
\label{sec:appendix:real_world_study_details}

\begin{table}[bt]
    \centering
    \caption{Survey Questionnaire for Realworld Developers}
    \label{tab:questionnaire}
    \begin{tabularx}{\linewidth}{@{} l X @{}} 
        \toprule
        \textbf{ID} & \textbf{Question} \\
        \midrule
        Q1 & How many years of software development experience do you have? \\
        \addlinespace
        Q2 & Have you heard of  business-logic vulnerabilities? \\
        \addlinespace
        Q3 & Have you dealt with business-logic vulnerabilities in your projects? \\
        \addlinespace
        Q4 & Based on your experience, how do you currently discover or test for business-logic vulnerabilities? \\
        \addlinespace
        Q5 & Do you think spending approximately one hour adding annotations in exchange for the potential to discover critical business logic vulnerabilities is worthwhile? \\
        \addlinespace
        Q6 & What do you see as the biggest potential obstacles to adopting such a tool in your project? \\
        \bottomrule
    \end{tabularx}
\end{table}

To supplement our student-based study and assess the real-world applicability of \tool, we conducted a qualitative study using questions listed in Table~\ref{tab:questionnaire} with 10 software developers and security analysts with various profession experience   (Q1). 
Among the participants, 1 had 1-3 years of experience, 4 had 3-5 years, 4 had 5-10 years, and 1 had more than 10 years. This study investigated their current practices, attitude of adopting \tool in their projects, and potential barriers to adoption.

We first confirmed the relevance of the problem domain (Q2+Q3). All participants reported that they were aware of and had personally encountered business-logic vulnerabilities in their projects. When asked about their current detection methods (Q4), the responses indicated a heavy reliance on manual and service-based processes. 
The most common methods were Manual Code Review (10/10) and External Penetration Testing (9/10). Other methods included Unit Testing (4/10) and Manual QA Testing (3/10). Notably, only one participant (1/10) had experience using static analysis tools for this purpose. This feedback confirms that detecting business-logic vulnerabilities remains a significant challenge that relies heavily on manual effort.
We then directly investigated the central trade-off of effort versus value. We asked participants if they would be willing to spend approximately one hour adding annotations to their project to help detect business-logic vulnerabilities (Q5). 
The response was unanimously positive, with a distinction based on project criticality. 4 participants stated they would "definitely use" \tool in their project. The remaining 6 participants reported they would apply it to applications based on their importance. 
This finding indicates that 100\% of the experienced participants see value in \tool and agree its annotation effort is a reasonable trade-off, particularly for projects having important logic.
Finally, we asked participants about their primary concerns regarding the adoption of \tool (Q6). The most significant barriers identified were: Learning Curve: All participants expressed concern about the time and effort required to learn a new annotation language. Integration Difficulty: 6/10 participants were concerned about \tool's compatibility and integration with their existing testing frameworks and CI/CD pipelines. Code Readability: 2/10 participants raised concerns that annotations could reduce code readability, especially without a standardized annotation style.

While the concern about the learning curve is valid, we note that our user study with students demonstrated that a brief training session was sufficient for participants to effectively use the \tool language. 
The other concerns regarding integration and standardization are practical implementation challenges that provide valuable feedback for developing a production-ready tool. We believe future work based on the \tool prototype can be refined through real-world usage and address this feedback.

In summary, this developer study confirms that business logic vulnerabilities are a prevalent and manually-tackled problem for experienced professionals. 
They perceive significant value in \tool's approach and, most importantly, view the required annotation effort as a reasonable trade-off for the potential security gains. 
The primary hurdles to adoption are practical engineering efforts, concerning the initial effort and workflow integration, which provides clear and valuable direction for future work on \tool's usability, tooling, and IDE/CI integration.

\section{Extended Discussion}
If a developer identifies a specific vulnerability during code review, they are likely to patch it directly rather than annotate it. However, this view assumes that vulnerabilities are obvious. In contrast, \tool targets subtle vulnerabilities that are easily overlooked, particularly those buried within complex implementations or opaque third-party libraries (as discussed in Appendix~\ref{sec:appendix:target_vuln}).
\tool's primary utility is to make implicit security intent machine-verifiable. In modern workflows, developers often treat APIs (e.g., \texttt{urlparse} in the motivation example List~\ref{lst:motivation_example}) as black boxes. 
They may understand a security requirement (e,g., block certain network requests) yet remain unaware of internal library flaws. 
By annotating this intent, \tool detects security vulnerabilities caused by API misuse or underlying bugs. 
This utility is validated by our discovery of zero-day vulnerabilities in audited projects like Home Assistant, enabling \tool to serve as a critical sanitizer where deep dependency introspection is infeasible.

\section{Performance Evaluation}
\label{sec:appendix:performance_eval}

The runtime performance measurement has two different setups. One is \emph{Trace All} that taints all the variables and the other is \emph{Monitor Syscall} that inspects all the system calls. 
Both setups are extreme setups because typical usages only need to inspect a subset of variables or system calls while executing the annotated code. 
The selected benchmark is the Python Performance Benchmark Suite~\cite{Pythonperfbench} provided by the CPython developers, which is intended to be an authoritative source of benchmarks for all Python implementations.
We run all the benchmarks in rigorous mode to get more accurate results, the evaluation result is shown in Table~\ref{Tab:perf}.
The \emph{Total} shows the name of each subtask.
\emph{Baseline} data is collected from an original, uninstrumented CPython interpreter, and the columns \emph{Monitor Syscall} and \emph{Trace All} are the overhead relative to the baseline execution.
As expected for our run-time performance optimization design choices, the average run-time performance overhead is relatively low, considering the benchmark is evaluated on extreme tracking-everything scenarios.
The overhead introduced by the system call monitor module is 5\% in average.
Among the test suites, \texttt{bench\_mp\_pool} and \texttt{bench\_thread\_pool} have a high overhead of around 35\% because the benchmark keeps using thread- and process-related system calls.
Similarly, \texttt{tornado\_http} has 44\% overhead because it sets up a server and invokes network-related system calls without other executions, so the overhead is much higher.
Comparing \texttt{logging\_silent} with \texttt{logging\_simple} and \texttt{logging\_format}, the latter have much higher overhead (29\% and 118\% compared to 1\%).
This is caused by writing log data to disk using the \texttt{write} system call, while \texttt{logging\_silent} keeps the log in memory without invoking system calls.
Our system call module inspects every byte written out by the \texttt{write} system call, hence the overhead is much higher.
For the \emph{Trace All} scenario, the average overhead is 10\%.
The test cases \texttt{sync\_tree\_none}, \texttt{async\_tree\_memoization}, \texttt{unpack\_sequence} and \texttt{bench\_mp\_poo} have more variables and byte code execution to track, so they have more overhead than other test cases.
Compared with Python dynamic analysis frameworks DynaPyt~\cite{eghbali2022dynapyt}, which incurs an overhead between $1.2\times$ and $16\times$ as evaluated by their paper, \tool has a much lower performance impact.
We attribute this to the instrumentation and tracking of \tool happening inside the interpreter, while DynaPyt's instrumentation is implemented as an external module in Python.

\begin{center}
\tablefirsthead{%
  \toprule
  {\bf Benchmark} & {\bf Baseline} & \multicolumn{2}{r}{\bf Monitor Syscall} & \multicolumn{2}{r}{\bf Trace All} \\
  \cmidrule(r){2-2} \cmidrule(r){3-4} \cmidrule(r){5-6}
  & {time $/ms$} & {time $/ms$} & $\Delta$ & {time $/ms$} & $\Delta$ \\
  \midrule
}
\tablehead{%
  \multicolumn{6}{l}{\textit{continued from previous page}} \\
  \midrule
}
\tabletail{%
  \midrule
  \multicolumn{6}{r}{\textit{continued on next page}} \\
}
\tablelasttail{\bottomrule}

\sisetup{
  table-format = 5.2,
  table-auto-round = true,
}
\tablecaption{Runtime Performance}\label{Tab:perf}
\scriptsize
\begin{supertabular}{l@{ \ }S@{ \ }S@{ \ }r@{ \ }S@{ \ }r}
{\bf Total} &15039.277443399998 &15721.4414742 & 5\% &16494.179284809 & 10\% \\
\addlinespace
2to3 &174 &188 & 8\% &189.66 & 9\% \\
async\_generators &197 &208 & 6\% &218.67 & 11\% \\
sync\_tree\_none &357 &382 & 7\% &431.97 & 21\% \\
async\_tree\_cpu\_io\_mixed &495 &532 & 7\% &598.95 & 21\% \\
async\_tree\_io &855 &935 & 9\% &974.7 & 14\% \\
async\_tree\_memoization &430 &451 & 5\% &533.2 & 24\% \\
asyncio\_tcp &360 &437 & 21\% &367.2 & 2\% \\
asyncio\_tcp\_ssl &1470 &1600 & 9\% &1484.7 & 1\% \\
asyncio\_websockets &163 &175 & 7\% &164.63 & 1\% \\
chameleon &4.94 &5.16 & 4\% &5.4834 & 11\% \\
chaos &57 &59.1 & 4\% &62.13 & 9\% \\
comprehensions &0.0155 &0.016 & 3\% &0.016895 & 9\% \\
bench\_mp\_pool &3.85 &5.11 & 33\% &4.697 & 22\% \\
bench\_thread\_pool &0.541 &0.752 & 39\% &0.60051 & 11\% \\
coroutines &15.5 &15.5 & 0\% &16.895 & 9\% \\
coverage &27.2 &27.5 & 1\% &29.104 & 7\% \\
crypto\_pyaes &62.9 &64.1 & 2\% &69.19 & 10\% \\
dask &212 &253 & 19\% &212 & 0\% \\
deepcopy &0.241 &0.252 & 5\% &0.26751 & 11\% \\
deepcopy\_reduce &0.00223 &0.00227 & 2\% &0.002453 & 10\% \\
deepcopy\_memo &0.0265 &0.027 & 2\% &0.02915 & 10\% \\
deltablue &3.87 &4.05 & 5\% &4.1022 & 6\% \\
django\_template &27.5 &28 & 2\% &29.975 & 9\% \\
docutils &1630 &1660 & 2\% &1939.7 & 19\% \\
dulwich\_log &38.9 &49.9 & 28\% &40.845 & 5\% \\
fannkuch &248 &249 & 0\% &280.24 & 13\% \\
float &56.3 &58.1 & 3\% &65.308 & 16\% \\
create\_gc\_cycles &0.756 &0.765 & 1\% &0.77112 & 2\% \\
gc\_traversal &1.83 &1.84 & 1\% &1.8483 & 1\% \\
generators &26.7 &26.7 & 0\% &27.234 & 2\% \\
genshi\_text &16.9 &17 & 1\% &18.928 & 12\% \\
genshi\_xml &32.7 &32.3 & -1\% &35.643 & 9\% \\
go &127 &127 & 0\% &138.43 & 9\% \\
hexiom &5.15 &5.12 & -1\% &5.562 & 8\% \\
html5lib &43.3 &44 & 2\% &45.465 & 5\% \\
json\_dumps &7.38 &7.49 & 1\% &7.8228 & 6\% \\
json\_loads &0.0128 &0.013 & 2\% &0.013568 & 6\% \\
logging\_format &0.00286 &0.00624 & 118\% &0.0031174 & 9\% \\
logging\_silent &0.0000973 &0.0000981 & 1\% &0.000104111 & 7\% \\
logging\_simple &0.00453 &0.00584 & 29\% &0.0049377 & 9\% \\
mako &8.35 &8.61 & 3\% &9.1015 & 9\% \\
mdp &1660 &1680 & 1\% &1693.2 & 2\% \\
meteor\_contest &63.8 &66.2 & 4\% &67.628 & 6\% \\
nbody &64.8 &68.1 & 5\% &78.408 & 21\% \\
nqueens &58.8 &59.7 & 2\% &62.328 & 6\% \\
pathlib &10.4 &17.5 & 68\% &11.024 & 6\% \\
pickle &0.00546 &0.00561 & 3\% &0.005733 & 5\% \\
pickle\_dict &0.0167 &0.0176 & 5\% &0.01837 & 10\% \\
pickle\_list &0.0023 &0.00231 & 0\% &0.002369 & 3\% \\
pickle\_pure\_python &0.245 &0.259 & 6\% &0.26705 & 9\% \\
pidigits &109 &114 & 5\% &110.09 & 1\% \\
pprint\_safe\_repr &567 &594 & 5\% &618.03 & 9\% \\
pprint\_pformat &1170 &1180 & 1\% &1275.3 & 9\% \\
pyflate &349 &355 & 2\% &380.41 & 9\% \\
python\_startup &5.49 &6.58 & 20\% &5.7096 & 4\% \\
python\_startup\_no\_site &3.56 &4.3 & 21\% &3.7024 & 4\% \\
raytrace &271 &277 & 2\% &287.26 & 6\% \\
regex\_compile &94.4 &95.1 & 1\% &102.896 & 9\% \\
regex\_dna &110 &112 & 2\% &113.3 & 3\% \\
regex\_effbot &1.52 &1.55 & 2\% &1.5808 & 4\% \\
regex\_v8 &12.9 &12.9 & 0\% &13.029 & 1\% \\
richards &41 &41.3 & 1\% &45.51 & 11\% \\
richards\_super &48.7 &51.1 & 5\% &52.109 & 7\% \\
scimark\_fft &191 &199 & 4\% &212.01 & 11\% \\
scimark\_lu &90.3 &93.4 & 3\% &96.621 & 7\% \\
scimark\_monte\_carlo &57.6 &58 & 1\% &63.36 & 10\% \\
scimark\_sor &105 &105 & 0\% &112.35 & 7\% \\
scimark\_sparse\_mat\_mult &3.04 &3.08 & 1\% &3.344 & 10\% \\
spectral\_norm &78.2 &78.1 & 0\% &86.02 & 10\% \\
sqlalchemy\_declarative &83.2 &86 & 3\% &88.192 & 6\% \\
sqlalchemy\_imperative &11 &11.1 & 1\% &11.44 & 4\% \\
sqlglot\_normalize &193 &196 & 2\% &206.51 & 7\% \\
sqlglot\_optimize &37.3 &37.7 & 1\% &40.657 & 9\% \\
sqlglot\_parse &1.1 &1.12 & 2\% &1.199 & 9\% \\
sqlglot\_transpile &1.32 &1.35 & 2\% &1.4256 & 8\% \\
sqlite\_synth &0.00152 &0.00153 & 1\% &0.0015504 & 2\% \\
sympy\_expand &308 &311 & 1\% &323.4 & 5\% \\
sympy\_integrate &13.6 &13.7 & 1\% &14.688 & 8\% \\
sympy\_sum &96.8 &97.8 & 1\% &102.608 & 6\% \\
sympy\_str &180 &183 & 2\% &194.4 & 8\% \\
telco &3.81 &3.86 & 1\% &4.0767 & 7\% \\
tomli\_loads &1490 &1520 & 2\% &1683.7 & 13\% \\
tornado\_http &69.6 &100 & 44\% &73.08 & 5\% \\
typing\_runtime\_protocols &0.313 &0.314 & 0\% &0.33178 & 6\% \\
unpack\_sequence &0.0000261 &0.0000261 & 0\% &0.000030798 & 18\% \\
unpickle &0.0076 &0.00764 & 1\% &0.00798 & 5\% \\
unpickle\_list &0.00232 &0.00231 & 0\% &0.0023664 & 2\% \\
unpickle\_pure\_python &0.171 &0.172 & 1\% &0.18639 & 9\% \\
xml\_etree\_parse &77 &77.1 & 0\% &81.62 & 6\% \\
xml\_etree\_iterparse &53 &54.6 & 3\% &56.18 & 6\% \\
xml\_etree\_generate &50.3 &51.9 & 3\% &53.318 & 6\% \\
xml\_etree\_process &43.1 &43.1 & 0\% &46.548 & 8\% \\
\end{supertabular}
\end{center}

\section{Skipped Applications}
\label{sec:appendix:skip_app}
As mentioned in Section~\ref{sec:eval_rediscover}, several packages were excluded from our experiment due to various limitations; these packages are listed in Table~\ref{tab:table_skip}. We skipped packages if: (1) we lacked the necessary domain knowledge for analysis (e.g., understanding the Qdrant API for qdrant-client or cryptographic algorithms for Cryptography); (2) we encountered unsatisfied requirements, often due to environment constraints (like lacking the required GPU setup for llama-cpp-python); (3) the package used multiple languages, including some not supported by our current prototype; or (4) the package functions primarily as a third-party library or plugin without clear entry points for the users (such as the pytest-cov testing plugin or the Transformers deep learning library).

\begin{table}[bt]
    \centering
    \sisetup{
        table-format=4.0,
        table-auto-round=true,
        drop-exponent=true,
        exponent-mode=fixed,
        fixed-exponent=3
    }
    \begin{threeparttable}
        \caption{Skipped Applications}
        \label{tab:table_skip}
        \scriptsize

        \begin{tabular}{@{ \ }r@{ \ }l@{ \ }S[table-format=3.1]@{ \ }l@{ }S[table-format=4.1]@{ \ }c@{ \ }}
            \toprule

            {\bf ID} & {\bf Name} & {\bf Stars} & {\bf Vuln. Identifier}  & {\bf LoC} & {\bf Reason} \\
                     &            & {$/ 10^3$}  &                         & {$/ 10^3$} &   \\
            \midrule

            1 & \href{https://github.com/qdrant/qdrant-client}{qdrant-client} & 918 & \textsc{CVE-2024-3829}& 67260 & Lack Knowledge \\
            2 & \href{https://github.com/Netflix/lemur}{lemur} & 1700 & \textsc{SNYK-7210298} &  37335 & Lack Knowledge \\
            3 & \href{https://github.com/NASA-AMMOS/AIT-Core}{AIT-Core} & 46 & \textsc{CVE-2024-35058} & 17790 & Lack Knowledge \\
            4 & \href{https://github.com/pyca/cryptography}{cryptography} &7000 & \textsc{CVE-2024-4603} & 56674 & Lack Knowledge \\
            5 & \href{https://github.com/abetlen/llama-cpp-python}{llama-cpp-python} & 8900 & \textsc{CVE-2024-34359} & 14405 & Missing HW Req.\\
            6 & \href{https://pypi.org/project/aliyundrive-webdav/}{aliyundrive-webdav} & 9700 & \textsc{CVE-2024-29640} & 162 & Unsupported Lang. \\
            7 & \href{https://github.com/pytest-dev/pytest-cov}{pytest-cov} & 1800 & \textsc{SNYK-6514860} & 2810 & Unclear Entry Pt. \\
            8 & \href{https://github.com/kjd/idna}{idna} & 256 & \textsc{SNYK-6597975} & 14459 & Lack Knowledge \\
            9 & \href{https://github.com/huggingface/transformers}{transformers} & 142000 & \textsc{CVE-2023-6730} & 1188188 & Unclear Entry Pt. \\
            10 & \href{https://github.com/MathItYT/manim-studio}{manim-studio} & 76500 & \textsc{SNYK-6141258} & 1450 & Lack Knowledge \\
            11 & \href{https://github.com/cryptoadvance/specter-desktop}{specter-desktop} & 823& \textsc{SNYK--6840403} & 39989 & Lack Knowledge\\
            12 & \href{https://github.com/keras-team/keras}{keras} & 62800 & \textsc{CVE-2024-3660} & 195486 & Lack Knowledge \\
            \bottomrule
        \end{tabular}
    \end{threeparttable}

\end{table}

\newpage
\begingroup
\centering
\tablecaption{Runtime overhead of Valgrind's MemCheck, and DynamoRIO's Memtrace}
\label{tab:dbi_perf}

\tablefirsthead{%
  \toprule
  {\bf Benchmark} & {\bf DynamoRio Memtrace} & {\bf Valgrind Memcheck}  \\
  \midrule
}
\tablehead{%
  \multicolumn{3}{l}{\textit{continued from previous page}} \\
  \midrule
}
\tabletail{%
  \midrule
  \multicolumn{3}{r}{\textit{continued on next page}} \\
}
\tablelasttail{\bottomrule}

\sisetup{
  table-format = 5.2,
  table-auto-round = true,
}
\scriptsize
\begin{supertabular}{lcc}

2to3 & 67.58x & 60.32x \\
async\_generators & 58.36x & 52.27x \\
async\_tree\_cpu\_io\_mixed & 38.20x & 49.46x \\
async\_tree\_io & 35.21x & 41.62x \\
async\_tree\_memoization & 41.47x & 49.05x \\
async\_tree\_none & 42.17x & 51.27x \\
asyncio\_tcp & 48.00x & 28.21x \\
asyncio\_tcp\_ssl & 67.76x & 117.77x \\
asyncio\_websockets & 24.87x & 23.09x \\
bench\_mp\_pool & 7.42x & 13.08x \\
bench\_thread\_pool & 10.31x & 11.11x \\
chameleon & 39.15x & 39.51x \\
chaos & 81.61x & 75.68x \\
comprehensions & 93.11x & 91.65x \\
coroutines & 91.68x & 57.53x \\
coverage & 47.21x & 29.66x \\
create\_gc\_cycles & 5.85x & 10.39x \\
crypto\_pyaes & 70.05x & 86.85x \\
deepcopy & 80.66x & 78.61x \\
deepcopy\_memo & 73.91x & 55.61x \\
deepcopy\_reduce & 81.87x & 86.15x \\
deltablue & 85.54x & 63.20x \\
django\_template & 82.05x & 79.49x \\
docutils & 48.75x & 54.29x \\
dulwich\_log & 45.73x & 56.80x \\
fannkuch & 76.17x & 60.84x \\
float & 63.45x & 58.90x \\
gc\_traversal & 13.18x & 23.28x \\
generators & 50.76x & 39.89x \\
genshi\_text & 80.53x & 73.58x \\
genshi\_xml & 73.35x & 68.65x \\
go & 76.82x & 61.44x \\
hexiom & 93.22x & 70.65x \\
html5lib & 56.77x & 57.91x \\
json\_dumps & 60.70x & 73.67x \\
json\_loads & 50.30x & 64.91x \\
logging\_format & 80.75x & 79.81x \\
logging\_silent & 76.86x & 41.93x \\
logging\_simple & 81.06x & 77.94x \\
mako & 68.19x & 65.13x \\
mdp & 50.07x & 61.03x \\
meteor\_contest & 47.84x & 45.03x \\
nbody & 69.54x & 42.85x \\
nqueens & 81.32x & 76.24x \\
pathlib & 51.38x & 72.21x \\
pickle & 56.46x & 57.44x \\
pickle\_dict & 28.73x & 37.78x \\
pickle\_list & 34.81x & 42.45x \\
pickle\_pure\_python & 85.54x & 80.77x \\
pidigits & 20.56x & 17.31x \\
pprint\_pformat & 76.60x & 86.75x \\
pprint\_safe\_repr & 76.64x & 85.50x \\
pyflate & 67.10x & 68.45x \\
python\_startup & 105.31x & 87.27x \\
python\_startup\_no\_site & 104.39x & 93.65x \\
raytrace & 84.78x & 72.12x \\
regex\_compile & 64.77x & 67.60x \\
regex\_dna & 14.98x & 21.82x \\
regex\_effbot & 26.74x & 54.05x \\
regex\_v8 & 35.75x & 39.93x \\
richards & 88.48x & 62.94x \\
richards\_super & 87.64x & 71.53x \\
scimark\_fft & 67.00x & 68.97x \\
scimark\_lu & 84.31x & 68.15x \\
scimark\_monte\_carlo & 84.31x & 81.36x \\
scimark\_sor & 91.47x & 82.54x \\
scimark\_sparse\_mat\_mult & 41.59x & 51.37x \\
spectral\_norm & 70.31x & 74.82x \\
sqlalchemy\_declarative & 43.19x & 48.89x \\
sqlalchemy\_imperative & 45.89x & 53.00x \\
sqlite\_synth & 85.72x & 93.92x \\
sympy\_expand & 65.94x & 73.24x \\
sympy\_integrate & 56.56x & 60.65x \\
sympy\_str & 62.08x & 69.54x \\
sympy\_sum & 55.87x & 66.06x \\
telco & 79.13x & 107.09x \\
tomli\_loads & 67.40x & 64.18x \\
tornado\_http & 34.50x & 37.21x \\
typing\_runtime\_protocols & 70.05x & 69.38x \\
unpack\_sequence & 21.81x & 21.68x \\
unpickle & 45.03x & 62.96x \\
unpickle\_list & 36.15x & 50.36x \\
unpickle\_pure\_python & 92.98x & 83.95x \\
xml\_etree\_generate & 78.27x & 84.38x \\
xml\_etree\_iterparse & 57.67x & 57.57x \\
xml\_etree\_parse & 47.09x & 57.45x \\
xml\_etree\_process & 78.95x & 80.96x \\
\end{supertabular}
\endgroup

\end{appendices}

\end{document}